\documentclass[aip,jcp,preprint]{revtex4-1}
\usepackage{amsfonts,amssymb,amsmath,latexsym}
\usepackage{xcolor}
\usepackage{setspace}
\usepackage{graphicx}
\usepackage{epstopdf} 
\usepackage[utf8]{inputenc} 
\usepackage[T1]{fontenc}

\usepackage{multirow}
\usepackage{tabularx}
\usepackage{float}
\usepackage{bigstrut}
\usepackage{booktabs}
 \usepackage[margin=0.80in]{geometry}
 \usepackage[normalem]{ulem}
 \newcommand{\stkout}[1]{\ifmmode\text{\sout{\ensuremath{#1}}}\else\sout{#1}\fi}

\newcommand{\elm}[3]{\left<#1\left|#2\right|#3\right>}

\newcommand{\lcpq}{Laboratoire de Chimie et Physique Quantiques, Universit\'e de Toulouse III Paul Sabatier, CNRS, 118 Route de Narbonne, F-31062, Toulouse, France}
\newcommand{\etsf}{European Theoretical Spectroscopy Facility (ETSF)}

\newcommand{\unife}{Dipartimento di Scienze Chimiche, Farmaceutiche ed Agrarie, Universit\`a di Ferrara, via Borsari 46, 44121 Ferrara, ITALY}

\begin{document}
\title{ 
Mapping of H\"uckel Zigzag Carbon Nanotubes onto independent Polyene chains:
application to periodic Nanotubes
}

\author{Gr\'egoire François}
\affiliation{\lcpq}

\author{Celestino Angeli}
\email{anc@unife.it}
\affiliation{\unife}

\author{Gian Luigi Bendazzoli}
\affiliation{Universit\`a di Bologna, via Irnerio 33, I-40126, Bologna, ITALY}

\author{V\'eronique Brumas}
\affiliation{\lcpq}

\author{Stefano Evangelisti}
\affiliation{\lcpq}

\author{J. Arjan Berger}
\affiliation{\lcpq}
\affiliation{\etsf}

\date{\today}

\begin{abstract}
{\bf Abstract:} 
The electric polarizability and the spread of the total position tensors are
used to characterize the metallic vs insulator nature of large (finite)
systems. Finite clusters are usually treated within the open boundary condition
formalism. This introduces border effects, which prevents a fast convergence to
the thermodynamic limit and which can be eliminated within the formalism of
periodic boundary conditions. Recently, we have introduced an original approach
to periodic boundary conditions, named Clifford Boundary Conditions. It
considers a finite fragment extracted from a periodic system and the
modification of its topology into that of a Clifford Torus.  The quantity
representing the position is modified in order to fulfill the system
periodicity. In this work, we apply the formalism of Clifford Boundary
Conditions to the case of Carbon Nanotubes, whose treatment results to be
particularly simple for the Zigzag geometry. Indeed, we demonstrate that at the
H\"uckel level these nanotubes, either finite or periodic, are formally equivalent to
a collection of {\em non-interacting dimerized linear chains}, thus simplifying their
treatment. This equivalence is used to describe some nanotube properties as the
sum of the contributions of the independent chains and to identify the origin
of peculiar behaviors (such as the conductivity). Indeed, if the number of
hexagons along the circumference is a multiple of three a metallic behavior is
found, namely a divergence of both the (per electron) polarizability and total
position spread of at least one linear chain. These results are in agreement
with those  in the literature from Tight-Binding calculations.

\end{abstract}

\maketitle

\section{Introduction}

\begin{figure}[ht]
\centering
\hspace{1cm}
\includegraphics[width=1.0\textwidth,angle=270]{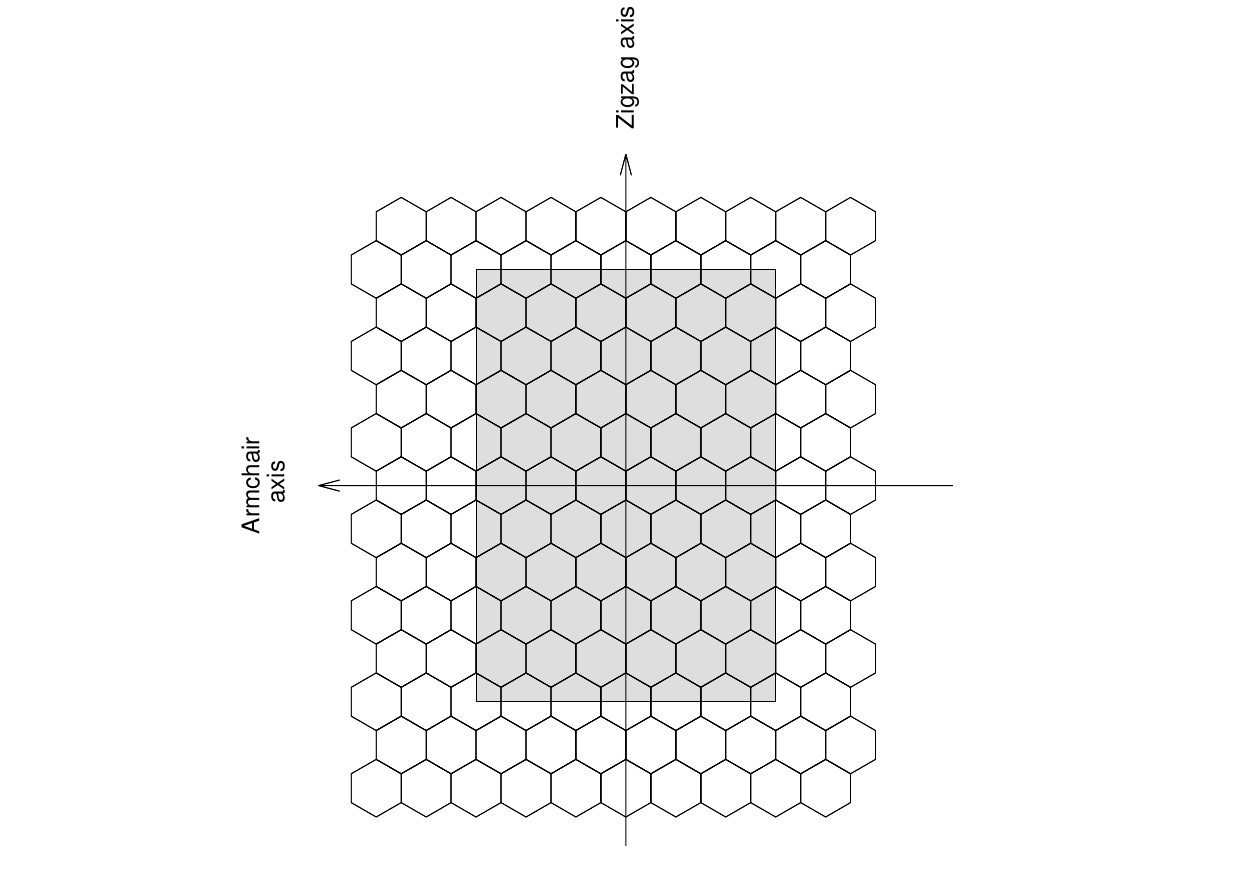}
\caption{Schematic representation of a graphene sheet and of the folding axes used to obtain a zigzag and an armchair nanotube.}
\label{fig:zigzag}
\end{figure}

Carbon Nanotubes were discovered in 1991 by Iijima,\cite{Iijima91} about fifteen years after the serendipitous discovery of Fullerenes, and have since then attracted much interest, both form the experimental and theoretical sides.
In particular, Single-Wall (SW) Nanotubes can be described as a  rectangular (in principle, of infinite length) sheet of graphene, folded in order to form a  cylinder. 
There are different types of nanotubes that we can derive from this description, changing length, width, and the angle between the sides of the graphene hexagons and of the sides of the rectangle. 
Ideal infinite nanotubes are characterized by two integer numbers, $(n,m)$, that correspond to the number of $n$ and $m$ unit vectors of the graphene lattice that are contained in
the so-called ``chiral vector''.
Nanotubes are generally classified into three different types. 
When the hexagons are oriented in such a way that a couple of opposite sides are parallel to the axis of the cylinder (of the nanotube), one has a zigzag nanotube $(n,0)$.
If a couple of opposite sides are orthogonal to the cylinder axis the nanotube is called an armchair nanotube $(n,n)$. 
Any other possible orientation in between is a nanotube with helicity (or a chiral nanotube). 
A schematic representation of the folding axes of a graphene sheet to obtain a zigzag and an armchair nanotube is reported in Fig. \ref{fig:zigzag}.

Each carbon atom in a nanotube has four valence electrons, three of which are
used for the $\sigma$-bonds, while the other electron is used to form
$\pi$-bonds (in these systems the $\sigma/\pi$ separation is not exact, see the
discussion in Sec. \ref{sec:formal}). The $\pi$ electrons can be much more
delocalized than their $\sigma$ counterpart and they are therefore responsible
of the conductivity when the nanotube is a conductor. For this reason, we will
focus hereafter on the $\pi$ part of the electronic structure making use of the
H\"uckel method to describe the system. From now on, we will focus only on
zigzag nanotubes of varying length and width.  Because we intend to work with
periodic boundary conditions we will model our nanotube as a flat rectangle of
graphene with periodic boundaries. It is worth noticing that this approach
opens the way to describe the properties of an infinite graphene sheet by
considering the limit of our results to the thermodynamic limit.  However, to
correctly be periodic the sheet of graphene that we choose must follow some
simple rules.  It needs to have an integer number of hexagons on the folded
dimension (the ``belt'' of the nanotube) and an even number of rings in the
direction of the nanotube axes.

It turns out that the conductivity of a Carbon Nano-Tube (CNT) is closely related to the associated integer $n$.
If $n$ is a multiple of three the energy spectrum of the nanotube is gapless, {\em i.e.}, the nanotube is a metal.
In the other cases, the nanotube is a semiconductor or an insulator.

In this article, we will focus our attention on zigzag nanotubes in the tight-binding, or H\"uckel, approximation.
In this case, it will be shown that, through a partial ({\em i.e.}, that mixes only atoms having a given longitudinal position along the tube) symmetry adaptation of the atomic orbitals of the system, a nanotube can be treated as a collection of non interacting polyene chains.
This is true regardless of the type of system, {\em i.e.}, both for finite CNTs in Open Boundary Conditions (OBC), and nanotubes treated within Periodic Boundary Conditions (PBC).
In order to investigate the conductivity properties of the nanotubes, we use the formalism of the Total-Position Spread (TPS), which is the second moment cumulant of the Total-Position operator.
In a series of works, Resta and coworkers\cite{Resta98, Resta99, RestaSorella, Resta02, Resta06} have shown that this quantity per electron (named also the Localization Tensor) diverges in the thermodynamic limit in the case of metallic systems, while it remains finite for insulators.
However, in the case of PBC, particular care must be used to deal with the position operator, since its usual definition is not compatible with the periodicity of the system.
For this reason, we recently proposed a modified definition of the position operator,\cite{Valenca,Angeli21,evangelisti_PhysRevB.105.235201} which can be used in the case of periodic systems and this modified version of the TPS  has been used in the present study.
In a similar way, we computed the per-electron electric polarizability of the system, that shows a behavior similar to TPS under the same metallic/insulator conditions.

This article is organized as follows:
In Section II we present the Position-Spread and Polarizability formalism, in the two different contexts, OBC and PBC.
In particular, for the PBC formalism we briefly recall the approach of Resta and coworkers,\cite{Resta98, Resta99, RestaSorella, Resta02, Resta06} and we present our alternative formulation.\cite{Valenca,Angeli21,evangelisti_PhysRevB.105.235201}
It has been shown elsewhere that our approach is identical to compute the two-dimensional trace of the TPS of a ring lying on a plane.\cite{Angeli21}
In Section III the formalism is applied to the particular case of zigzag nanotubes treated at the tight-binding (H\"uckel) level.
In Section IV it is shown how a zigzag nanotube can be formally seen as a set of non-interacting dimerized polyenes in the case of OBC or of non-interacting annulenes in the case of PBC. 
Such a modeling is independent of the type of boundary conditions, although, for the sake of simplicity, only PBC nanotubes will be considered here.
This result opens the possibility of the treatment of very large nanotubes and to describe the behaviour of the nanotube as the sum of the behavior of the individual one-dimensional chains, for which the analysis is much easier. 
In particular when PBC are used, the H\"uckel model gives analytic solutions for the dimerized chains and the analysis of their properties is particularly simple. 
Finally, in Section V, some conclusions are drawn. 

\section{Formalism}
\label{sec:II}

\subsection{Open and Periodic Boundary Conditions}

We remind here briefly the formalism that we recently
developed\cite{Tavernier_2020,Tavernier_2021,Alves21,Azor21} for the treatment
of infinite periodic systems, {\em i.e.,} systems composed of an infinite
number of repeated unit cells. This formalism has shown to be
effective in a few application.\cite{Gong21,Yagi22,Chen23} Our approach is
based on three main ideas: 1) the extraction of a finite fragment out of the
infinite system; 2) the modification of the topology of the fragment to that of
a Clifford torus;\cite{Cliff71,Klein90,Bianchi96,Volke13,McInt99,Schwa11} and
3) the definition of a modified position (or position operator, in the case of
quantum systems) on the Clifford torus.  To this purpose, we first define a
supercell, which is a parallelepiped subset of unit cells of the entire system.
Then, in order to avoid border effects, the topology of the supercell is
modified into the topology of a  Clifford torus.  A Clifford torus is a {\em
flat borderless} $d$-dimensional manifold, obtained by joining the opposite
faces of a $d$-dimensional parallelepiped without deformation (see for instance
Ref. [\onlinecite{Schwar11}]).  It can be embedded, without distortion, in an
ordinary $2d$ space (either $\mathbb{R}^{2d}$ or, equivalently,
$\mathbb{C}^d$).  If $d=1$, it is isomorphic to a circle (in $\mathbb{R}^2$ or
$\mathbb{C}^1$), while for $d>1$ it cannot be represented without distortions
in our usual 3D space.

In order to treat periodicity, a crucial element is the definition of the
distance between two points on the torus.  In this work, we deal with
nanotubes, which are quasi-one-dimensional systems and only the periodicity in
the direction of the tube axis has to be taken into account.  Therefore, only
the longitudinal position must be defined, in such a way to be able to compute
the longitudinal position spread and polarizability.  In our formalism, the
periodic complex position $q(z)$ of a point having longitudinal coordinate $z$
is defined as
\begin{equation}
q(z) = \frac{L}{2\pi i} \left[ \exp\left({i\frac{2\pi}{L}{z}}\right) -1 \right] \hspace{6mm} .
\label{eq:complexpos1D}
\end{equation}
where $L$ is the length of the supercell.
Notice that $q(z)$ is a complex periodic function of $z$, having a period equal to $L$.
In the formalism of quantum mechanics, a position operator must be introduced.
It is given by the multiplication by the position reported in Eq. (\ref{eq:complexpos1D}), in a completely analogous way to what is done with the ordinary position operator.

The distance between two points $z_1$ and $z_2$ can also be defined and it is
given by the modulus of the difference between the periodic positions of the
two points:
\begin{equation}
    d(z_1,z_2) = |q(z_1)-q(z_2)|
\end{equation}
which is a real quantity, as it should.  We note that this distance does not
correspond to the geodesic Riemannian distance (on the torus) between two
points.  It is rather the Euclidean distance between the two point, taken in
the space in which the torus is embedded.\cite{Tavernier_2021} This definition
has the advantage that the distance becomes a continuous differentiable
function of the coordinates of the points, a property not shared by the
geodesic distance.  This property is particularly important for the treatment
of the Coulomb interaction between charged particles, although this interaction
is not taken into account in the Hückel Hamiltonian, the operator here
considered.  However, it plays also an important role in the definition of the
total position spread and polarizability, two quantities studied in the present
work.

Finally, we want to stress the fact that the symmetry treatment here proposed
for zigzag carbon nanotubes (see Sec. \ref{sec:formal}) is completely general
and not associated to the periodicity of the tubes.  Indeed, it can be applied
both to infinite periodic and finite systems.

\subsection{The Position Spread}

The Total-Position Spread (TPS), introduced by Resta some time
ago,\cite{Resta98, Resta99, RestaSorella, Resta02, Resta06} plays a very
important role in the Electronic Structure Theory.  It measures the mobility of
the electrons in molecular or extended systems, and is therefore capable of
distinguishing conductors from insulators.  In particular, we have applied in
the past this formalism to describe the bond structure in molecular
systems.\cite{Angeli13, Brea13, Benda14, Khatib14b,Brea16, Segal18, Chagl19}
However, this formalism is most interesting in the case of the treatment of
extended systems.\cite{Vetere08, Monari08, Monari08a, Vetere09, Evang2010,
Benda10, Benda11, Benda12, Giner13, Benda15, Khatib15, Ferti15, Diaz18,
Huran18, Valenca, Angeli21}

In a system subjected to OBC, the total position operator, from which the total position spread (TPS) is constructed,
is the one-body operator defined as the sum over the individual position operators of the various electrons,
\begin{equation}
\hat{ X} \; = \; \sum_\mu \, \hat{ x}_\mu
\label{eq:posoper}
\end{equation}
where $\hat{x}_\mu$ is the position operator of electron $\mu$.
The total position spread $ \Lambda$ is a real quantity, defined
as the second cumulant moment of the complex total position operator as
\begin{equation}
{ \Lambda} \; = \; \langle \Psi|\hat{ X}^2 |\Psi\rangle \, - \langle \Psi|\hat{ X} |\Psi\rangle ^2.
\label{eq:lambaR}
\end{equation}
The TPS is translationally invariant and its trace is a constant.
In a complete analogy with the OBC definition, in a PBC context we replace in Eq. \ref{eq:posoper} the position of a particle, $\hat{ x}_\mu$, by its complex 
position, $\hat{ q}_\mu$. Indeed,
the problem with PBC is that the standard definition of the position is not compatible with the periodicity of the system.\cite{PhysRevLett.80.1800}
In such a way, the complex total position operator is still a one-body operator, defined as
\begin{equation}
\hat{ Q} \; = \; \sum_\mu \, \hat{ q}_\mu
\end{equation}
where $\hat{ q}_\mu$ is the complex position operator of electron $\mu$, which, following Eq. \ref{eq:complexpos1D}, is given by
\begin{equation}
\hat { q}_\mu= \frac{L}{2\pi i} \left[\exp\left({i\frac{2\pi}{L}\hat{ x}_\mu}\right) -1 \right].
\end{equation}

The TPS $ \Lambda$ within PBC is still a real quantity, defined
as the second cumulant moment of the complex total position operator,
\begin{equation}
{ \Lambda} \; = \; \langle \Psi_0|\hat{ Q}^{\dagger} \, \hat{ Q} |\Psi_0\rangle \, - \, \langle \Psi_0|\hat{ Q}^{\dagger}|\Psi_0\rangle \langle\Psi_0|\hat{ Q} |\Psi_0\rangle
\end{equation}
in a complete analogy with the OBC definition. We use here the notation $\Psi_0$ for the wave function of the state under consideration (usually, the ground state) to make more clear the following derivation.
In order to compute $ \Lambda$, it is convenient to make use of  the resolution of the identity operator $\hat 1$, which is given by $\hat 1 \, = \, \sum_I |\Psi_I \rangle \langle \Psi_I|$, where $\left\{|\Psi_I\rangle\right\}$ is an orthonormal basis set of the Hilbert space, containing $|\Psi_0\rangle$ as an element.
The resolution of the identity is inserted between the operators $ \hat Q^+$ and $ \hat Q$, obtaining
\begin{equation}
{ \Lambda} \; = \; \sum_{I\neq 0} \, \langle \Psi_0|\hat{ Q}^{\dagger} \, |\Psi_I \rangle \langle \Psi_I| \, \hat{ Q} |\Psi_0\rangle.
\label{eq:lambda1}
\end{equation}
If the $|\Psi_I\rangle$ are the exact eigenvectors of Hamiltonian (supposed to be known) this equation is particularly suitable to perform numerical calculations. If only approximate eigenvectors are available, Eq. \ref{eq:lambda1} is still suitable for computation.

\subsection{The Polarizability}

The electric polarizability, ${ \Pi}$, in the OBC formalism, is given by
\begin{equation}
{ \Pi} \; = \; \langle \Psi_0|\hat{ R}^{\dagger} \, (\hat H-E_0)^{-1}_\perp \, \hat{ R} |\Psi_0\rangle 
\end{equation}
where the operator $(\hat H_0-E_0)_\perp$ is the restriction of $(\hat H_0-E_0)$ to the orthogonal complement of the ground state $|\Psi_0\rangle$.
In a similar way, we can define the polarizability $ \Pi$ within PBC as the real quantity
\begin{equation}
{ \Pi} \; = \; \langle \Psi_0|\hat{ Q}^{\dagger} \, (\hat H-E_0)^{-1}_\perp \, \hat{ Q} |\Psi_0\rangle.
\end{equation}
In analogy with the TPS case, we insert now two resolutions of the identity in the previous equation, and we get
\begin{equation}
{ \Pi} \; = \; \sum_{I\neq 0}\sum_{J \neq 0} \, \langle \Psi_0|\hat{ Q}^{\dagger} \, |\Psi_I \rangle \langle \Psi_I| \, (\hat H-E_0)^{-1}_\perp \, |\Psi_J \rangle \langle \Psi_J| \,\hat{ Q} |\Psi_0\rangle.
\end{equation}
Supposing also in this case that the $|\Psi_I\rangle$ are the eigenvectors of the Hamiltonian, 
this equation can be conveniently rewritten as
\begin{equation}
{ \Pi} \; = \; \sum_{I \neq 0} \, \frac{\langle \Psi_0|\hat{ Q}^{\dagger} \, |\Psi_I \rangle\langle \Psi_I| \,\hat{ Q} |\Psi_0\rangle}{E_I-E_0} \, .
\end{equation}
Again, as for the TPS case, this is a form particularly adapted to perform numerical calculations, provided the eigenvectors $|\Psi_I\rangle$ of the Hamiltonian are known. 
In the next section, these equations will be applied to the particular case of the H\"uckel Hamiltonian. 

\section{Polarizability and Total Position Spread
for one-electron Hamiltonians
\label{sec:tpspol}}

In this section, we apply to a nanotube described with the H\"uckel Hamiltonian
the formalism obtained in Sec. \ref{sec:II} for the polarizability and the TPS
in the case of periodic systems.  The calculation is much simplified by the
fact that, for a one-electron
Hamiltonian, the Slater determinants, such as for instance the singly-excited
determinants from $|\Psi_0\rangle$, are the exact eigenvectors of the
Hamiltonian.  Because of the one-electron nature of the position operator, both
in OBC and PBC, these single excitations are the only determinants in the
resolution of the identity giving a non vanishing contribution.  In the
following we use the notation $|\Phi\rangle$ to indicate Slater determinants.

\subsection{Periodic Boundary Conditions: Total Position Spread\label{sec:tps}}

We consider a system described by $N$ spinorbitals.
For the TPS, we get
\begin{equation}
    { \Lambda} \; = \; \sum_{j \in {\rm O}} \sum_{l \in {\rm V}} 
    \langle\Phi_0|{ \hat Q^+}|\Phi_j^l\rangle \langle\Phi_j^l|{ \hat Q} |\Phi_0\rangle, %  \;\;\; ,
\end{equation}
where $\rm O$ and $\rm V$ represent the sets of occupied and virtual spinorbitals, respectively, and $|\Phi_j^l\rangle$ are singly excited determinants with respect to $|\Phi_0\rangle$ (in $|\Phi_0\rangle$, spinorbital $j$ has been replaced by spinorbital $l$).
By using the Slater Rules, this equation becomes
\begin{equation}
    { \Lambda} \; = \; \sum_{j \in {\rm O}} \sum_{l \in {\rm V}} 
    \langle\phi_j|{ \hat q}^+ |\phi_l\rangle \langle\phi_l|{ \hat q} |\phi_j\rangle.
\end{equation}
By taking into account the expression of the $\phi_j$ MOs in terms of the $\chi_\rho$ AOs,
\begin{equation}
    \phi_j=\sum_{\rho=1}^N c_{\rho,j} \chi_\rho,
\end{equation}
we obtain for the TPS the expression in the AO basis
\begin{equation}
    { \Lambda} \; = \; \sum_{j \in {\rm O}} \sum_{l \in {\rm V}} \; \Big| \sum_{\rho,\sigma =1}^N c_{\rho,j} \, c_{\sigma,l} \;
    \langle\chi_\rho|{ \hat q} |\chi_\sigma\rangle \Big|^2.% \;\;\; .
\end{equation}

\subsection{Periodic Boundary Conditions: Polarizability\label{sec:polar}}

For the Polarizability $ \Pi$, one has 
\begin{equation}
   { \Pi} = \; \sum_{j \in {\rm O}} \sum_{l \in {\rm V}} 
    \langle\Phi_0|{ \hat Q^+}|\Phi_j^l\rangle \;  \frac{1}{\epsilon_l-\epsilon_j} \; \langle\Phi_j^l|{ \hat Q} |\Phi_0\rangle.
\end{equation}
By using a procedure similar to that applied for the total position spread, we obtain the final expression 
\begin{equation}
    { \Pi} \; = \; \sum_{j \in {\rm O}} \sum_{l \in {\rm V}} \; \frac{\Big| \sum_{\rho,\sigma =1}^N c_{\rho,j} \, c_{\sigma,l} \;
    \langle\chi_\rho|{ \hat q} |\chi_\sigma\rangle \Big|^2}{\epsilon_l-\epsilon_j  }.
\end{equation}

\section{Zigzag Carbon Nanotubes: Formal Treatment\label{sec:formal}}

\begin{figure}[ht]
\centering
\hspace{1cm}
\includegraphics[width=0.9\textwidth]{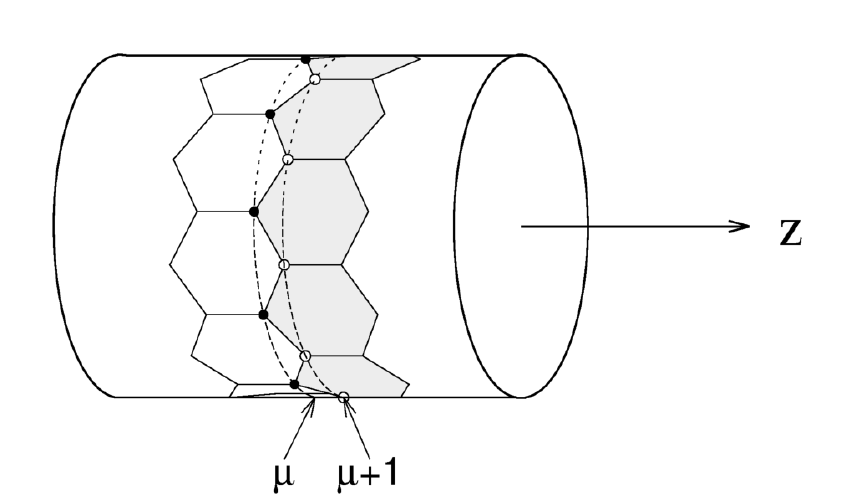}
\caption{Schematic representation of a zigzag nanotube. The atomic system lying on the line shared by two hexagon belts (in gray and white, respectively) is used in the text as an ``annulene'' and contains two sets of carbon atoms (indicated with empty and full circles, respectively). 
The atoms belonging the same set have the same coordinate on the $z$ axis, while the two sets refer to different $z$ coordinates. The two sets are indicated in figure with $\mu$ 
and $\mu+1$.}
\label{fig:zig_num}
\end{figure}

In order to define the formalism used to describe a nanotube, we consider the case of a fragment of a zigzag nanotube subjected to PBC.
As commonly done, the nanotube axis is chosen to be the $z$ axis. 
A zigzag nanotube can be seen as a collection of $m$ interacting all-trans annulenes, each one consisting of $2n$ Carbon atoms that follow the zigzag path going around the tube.
The number of carbon atoms in each annulene must be even, in order to form the $n$ hexagons that lie on the tube (in gray in Fig. \ref{fig:zig_num} the sets of $n$ hexagons).
In order to apply PBC, and to be able to match the two extremities of the tube without the need of a twist, the number $m$ must be even.
Notice that the analysis that follows can be easily generalized to treat also the case of open-ended nanotubes.
In this case, however, no constraint has to be imposed on the parity of $m$.
Because of the all-trans geometry of the annulenes, two types of carbon atoms are present in every cycle, each type being characterized by a common value of the $z$ coordinate.
In a nanotube, the Hydrogen atoms on each annulene are obviously absent, and replaced by bonds with neighbour annulenes, except for the two edges at the nanotube extremities in the OBC case.  

With this notation, a nanotube contains a total of $N=2mn$ Carbon atoms. 
For the formalism developed in this work, it is crucial to note that the Carbon atoms can be grouped in $2m$ sets of $n$ atoms, two sets in each ring, and each set containing those atoms that share the same $z$ coordinate. 
The atoms belonging to the same set are equivalent by symmetry and can be transformed into each other by one of the symmetry operations of the $C_n$ symmetry point group (the group containing all rotations around the $z$ axis of a multiple of a $2\pi/n$ angle). 
Each annulene is composed of two of these sets (there are therefore a total of $2m$ sets in the whole tube) and moving along each annulene the Carbon atoms belong to one or to the other set, alternately. The index $\mu$ is used hereafter to enumerate the different Carbon-atom sets, ($1\leq\mu\leq 2m$). 
For the sake of clarity, a schematic representation of a zigzag nanotube is reported in Fig. \ref{fig:zig_num}, with a representation of one annulene and of the two sets of Carbon atoms belonging to it.
Four different types of Carbon atom sets can be identified depending on the value $\mod{(\mu,4})$. 
Note that in periodic nanotubes, since $m$ must be even, $2m$ is necessarily a multiple of four. 
The Carbon atoms belonging to a given set are identified by using the index $\nu$, with ($0\leq\nu\leq n-1$).

In planar conjugated organic hydrocarbons, one can rigorously divide the molecular orbitals into two groups: the $\sigma$ MOs (\textit{even} with respect to the reflection in the molecular plane), and the $\pi$ MOs (\textit{odd}). 
Such a division is not strictly valid for nanotubes, since they are not flat structures, but the conceptual separation between $\sigma$ and $\pi$ orbitals can still be kept, at least at an approximate level.
In this case, the nanotube surface plays the role of the molecular
plane. 
In this work, the focus is on the (pseudo) $\pi$ MOs, which are linear combinations of the radial $p$ AOs (those that are orthogonal to the nanotube surface, which actually have a small component on the $2s$ Carbon AOs.), one per Carbon atom. 
These atomic orbital are labelled as $p_{\mu,\nu}$, where the atomic indices $\mu$ and $\nu$ have been defined above.

A zigzag nanotube belongs to the $C_n$ symmetry point group, whose
irreducible representations (irreps) can be labeled by an index $j$ ($0\leq j\leq n-1$) and the character in the $j$-th irrep of the $C_n^l$ element of the group (a rotation of an angle of $ 2\pi l /n$ around the $z$ axis, $0\leq l \leq n-1$) is $\exp{\left(i\frac{jl 2\pi}{n}\right)}$. 

The $p_{\mu,\nu}$  AOs sharing the same index $\mu$ can be linearly combined to obtain symmetry-adapted basis functions $\phi_{\mu,j}$ (carrying the $j$ irrep), by applying the projector on the $j$-th irrep, $\hat{P}_j$ 
\begin{equation}
    \hat{P}_j=\frac{1}{\sqrt{n}}\sum_{l=0}^{n-1}e^{i\frac{j l 2\pi}{n}}C_n^l
\end{equation}
to any given $p_{\mu, \nu}$, \textit{e.g.} $p_{\mu,0}$ :
\begin{equation}
    \phi_{\mu,j} = \hat{P}_j \; p_{\mu,0}= \frac{1}{\sqrt{n}}\sum_{\nu=0}^{n-1} e^{i\frac{j\nu2\pi}{n}} p_{\mu, \nu} \; . 
    \label{eq:defphijnu}
\end{equation}
These Symmetry-Adapted Linear Combinations (SALC) of AOs  are the basis set used to compute the MOs. Within this approach, $\phi_{\mu,j}$ can be seen as the $j$\textit{-th} function lying on the $\mu$\textit{-th} ``site'' of the system (we use this term for the sake of simplicity, but, actually, one should use the more appropriate term ``pseudo-site''). In this frame, the $\mu-th$ site has a well defined $z$ coordinate, while the $x$ and $y$ coordinates are not defined. Notwithstanding the ambiguity concerning its position the three-dimensional space,  the concept of ``site'' is used hereafter, given that for nanotubes the main focus is on the $z$-dependent properties (nanotubes are often considered as quasi-1D systems.

Within the H\"uckel approximation, the use of the $\{\phi_{\mu,j}\}$ basis set for all $\mu$  makes, in general, the matrix elements of the one electron Hamiltonian $\hat{h}$ complex. 
We recall that in this approximation the  matrix elements of $\hat{h}$ on the $p$ atomic orbitals are different from zero only between AOs lying on atoms directly connected. 
The off-diagonal non-zero interactions are all equal, and are usually indicated in quantum chemistry with the symbol $\beta$ (and called resonance integral), while in solid-state physics this parameter is known as hopping integral and denoted by the symbol $t$ ($t=-\beta$). 
Notice that in most physical systems $t$ is positive, and hence $\beta$ is negative.
Our application of the H\"uckel approach
to CNTs introduces additional approximation with respect to the standard H\"uckel method. Indeed, the fact that the Carbon atoms are not on a plane makes in principle the energy of the $p_{\mu, \nu}$ AOs dependent on the curvature of the CNT surface (and thus on $n$). Moreover, the axis of the $p_{\mu, \nu}$ AOs are not all parallel. The dihedral angle between nearest neighbor $p_{\mu, \nu}$ AOs (when it is different from zero) depends on the curvature and also affects in principle the value of the $\beta$ parameter between them (thus different $\beta$ values are present in the same nanotube and different nanotubes are described with different $\beta$ values). The dependence of the H\"uckel parameters on $n$ allows to study the energy of bending  deformations of graphene.\cite{Zhang11,Nikif14} To keep the treatment
simple, in the present work this dependence is ignored. The full inclusion of these subtle effects will be explored in future works.

Even if the appearance of complex matrix elements is not a strong complication, we prefer here to introduce a simple modification of the definition of the symmetry adapted functions in Eq. \ref{eq:defphijnu} which ensures that all matrix elements of $\hat{h}$ are real. 
In this approach, Eq. \ref{eq:defphijnu} is kept when $\mod{(\mu,4)}=0$ or $1$ (sites of type 0 or 1), while for $\mod{(\mu,4)}=2$ or $3$ (sites of type 2 or 3) the basis functions are multiplied by the $j$-dependent phase factor $\exp{(i\frac{j\pi}{n})}$:
\begin{equation}
    \phi_{\mu,j} \; = 
    e^{i\frac{j\pi}{n}}\frac{1}{\sqrt{n}} \, \sum_{\nu=0}^{n-1} e^{i\frac{j\nu 2\pi}{n}} p_{\mu, \nu} =
    \; \frac{1}{\sqrt{n}} \, \sum_{\nu=0}^{n-1} e^{i\frac{j(\nu+1/2)2\pi}{n}} p_{\mu, \nu}.
     \label{eq:site23}
\end{equation}

It is worth noticing that with this choice we have taken benefit of a degree of
freedom in the definition of any function in quantum mechanics and that this
choice does not change the final results, both for the energies and for the
wave functions and therefore for any property derived from the wave functions
(as the TPS or the polarizability). 

Concerning the matrix elements of $\hat{h}$ on the so defined symmetry adapted
basis, we first notice that the elements between functions sharing the same
$\mu$ (among which one has the diagonal elements) are vanishing since the
$p_{\mu,\nu}$ AOs sharing the same $\mu$ value lies on Carbon atoms which are
not connected. The same is true for two basis functions with $\mu$ differing by
more than one, again because the corresponding Carbon atoms are not connected.
One has therefore
\begin{equation}
    \elm{\phi_{\mu,j}}{\hat{h}}{\phi_{\mu^\prime,j^\prime}}=0~~~~\forall j,j^\prime~~~\mathrm{if}~\mu=\mu^\prime
    ~\mathrm{or}~\left|\mu-\mu^\prime\right|> 1.
    \label{eq:fock1}
\end{equation}

Focusing on the case $\mu^\prime=\mu+1$, one has to consider two possibilities:
\begin{enumerate}
    \item $\mod{(\mu,4)}=0$ or $2$
\begin{equation}
    \elm{\phi_{\mu,j}}{\hat{h}}{\phi_{\mu+1,j^\prime}}=\delta_{j,j^\prime}\;\beta;
    \label{eq:fock2}
\end{equation}
\item $\mod{(\mu,4)}=1$ or $3$
\begin{equation}
    \elm{\phi_{\mu,j}}{\hat{h}}{\phi_{\mu+1,j^\prime}}=\delta_{j,j^\prime}\;2\beta \cos\left(\frac{j\pi}{n}\right).
    \label{eq:fock3}
\end{equation}
\end{enumerate}

In the case of a periodic nanotube an additional case must be added:
\begin{equation}
    \elm{\phi_{1,j}}{\hat{h}}{\phi_{2m,j^\prime}}=\delta_{j,j^\prime}\;\beta.
    \label{eq:fock4}
\end{equation}
The results reported in Eqs. \ref{eq:fock1}-\ref{eq:fock4} show that in the
symmetry adapted basis set $\left\{\phi_{\mu,j}\right\}$ (conveniently
organized) the $\hat{h}$ matrix is zero on its diagonal. Furthermore, it is
block diagonal with each of the $n$ square blocks being a tridiagonal matrix
(in case of open-ended nanotubes, two more non zero elements in each block for
periodic nanotube) with dimension $2m \times 2m$, each block corresponding to a
given $j$ value.

As a result of such a Hamiltonian-matrix  structure, we can state that an open
zigzag nanotube is equivalent to a set of $n$ independent, linear, and possibly
dimerized, chains containing $2m$ sites (a site for each $\mu$ value carrying a
single function $\phi_{\mu,j}$), one chain for each value of $j$.  Within PBC,
the linear (dimerized) chains are folded, in order to form a (dimerized)
annulene.  To avoid confusion we stress again that the independent linear
chains (each one defined by the index $j$) are not formed by real sites (Carbon
atoms), but rather by fictitious sites (with a well defined position on the $z$
coordinate depending on $\mu$) bearing the $\phi_{\mu,j}$ functions, which are
linear combination of $p_{\mu,\nu}$ AOs.

The energy of these site functions is vanishing, while the interactions are
restricted to next neighbour sites (as in the 1D Carbon atom chains) and
assumes values which alternates between an effective value depending on $j$,
$\beta_{\rm e}(j)=2\beta \cos\left({j\pi}/{n}\right)$, and the standard
interaction defined in the H\"uckel approximation, $\beta$. 

A few simple observations can be inferred from these considerations:
\begin{itemize}
    \item  when $n$ is even, in the chain corresponding to $j=n/2$ the interaction $\beta_{\rm e}(j)$ is vanishing.  Therefore, with PBC this chain contains $m$ isolated couples of sites with a $\beta$ interaction among themselves (behaving as $m$ isolated ethylene molecules) and one has therefore $m$ MOs at an energy $\beta$ and $m$ MOs at $-\beta$. On the other hand with OBC the first and last sites of the chain are isolated (giving rise to two edge states at zero energy, where the two isolated sites are singly occupied), while the rest of the chain is composed of $m-1$ isolated couples of sites with a $\beta$ interaction among themselves (the degeneracy of the $\beta$ and $-\beta$ orbital energy is therefore $m-1$).
    
    \item when $\mod{(n,3)}=0$, in the chains corresponding to $j=n/3$ and $j=2n/3$ the interaction $\beta_{\rm e}(j)$ assumes the value $\beta$ and $-\beta$, respectively. These chains are therefore  non-dimerized chains. 
    It is well known (see, for instance, Ref. [\onlinecite{Monari08}]) that a non-dimerized chain is a conductor, and therefore also the whole nanotube is a conductor (see also the comment after Eq. \ref{eq:defepsilon}). 
\end{itemize}
Focusing now on a nanotube with PBCs, we notice that the chains are composed of $m$ unit cells, having two sites per unit cell.
As discussed in Ref. [\onlinecite{Angeli21}] (see also Appendix A), only the $2\times 2$ building blocks on the main diagonal and the block immediately adjacent to them are non vanishing. 
In particular, the diagonal blocks of the Hamiltonian for each chain are constant, and are given by
\begin{equation}
{{\mathrm{\mathbf H}}}_0 \; = \;
\left\|
\begin{array}{ccccccccc}
0     & \beta \\
\beta & 0     \\
\end{array}
\right\|  \hspace{6mm} ,
\end{equation}
while the only non vanishing off-diagonal block is ${{\mathrm{\mathbf H}}}_1$, which depends on the chain index,
\begin{equation}
{{\mathrm{\mathbf H}}}_1 (j) \; = \;
\left\|
\begin{array}{ccccccccc}
0                & 0     \\
\beta_{e}(j) & 0     \\
\end{array}
\right\| \hspace{10mm} j=0,\ldots, n-1.
\end{equation}
Therefore, the Effective Hamiltonian depends in this case on $k$ and $j$ and it can be written as
\begin{equation}
{\mathrm{\mathbf{\bar H}}}(k,j) \; = \;
\left\|
\begin{array}{ccccccccc}
0     & \beta + \beta_{e}(j) \, e^{-ik\theta}\\
\beta + \beta_{e}(j) \, e^{ik\theta}& 0     \\
\end{array}
\right\| \hspace{10mm} 
\begin{array}{c}
j=0,\ldots, n-1\\
k=0,\ldots, m-1
\end{array}\hspace{6mm} .
\end{equation}

It is worth noticing that using PBC, it is also possible to use
the symmetry of the Hamiltonian in order to obtain analytical  solutions for each $j$.
The eigenvalues form two symmetric bands, given by
\begin{equation}
\varepsilon_\pm^j(k) = \pm \sqrt{\beta^2+\beta_{e}^2(j)+2\beta\beta_{e}(j)\cos k\theta}
\label{eq:defepsilon}
\end{equation}
where $k=0,1,...,m-1$, and $\theta=2\pi/m$. 
If, for some value of $j$, the effective interaction $\beta_e$ is equal to $\beta$, the previous equation (\ref{eq:defepsilon}) becomes 
\begin{equation}
\varepsilon^j_\pm(k) = \pm \beta\sqrt{2 \, (1+\cos {k\theta})} = \pm 2 \beta \cos{k\theta/2},
\label{eq:defepsilon_0}
\end{equation}
which are the energy bands of an undimerized chain.
In this case, for $\cos(k\theta)=-1$, the argument of the square root is zero, and we have $\varepsilon_\pm(k) = 0$.
This happens for $k\theta =  \pi$ (that is, $k=m/2$): in this case the two bands $\varepsilon_\pm^j$ have zero gap at the Fermi level, and the two corresponding chains are metallic.

This means that the whole nanotube is also a conductor.
On the other hand, if $\beta_{e}(j) \ne \beta$, the argument of the square root in Eq. (\ref{eq:defepsilon}) is always positive, a gap is developed around the Fermi level and the chain is an insulator.
The only possibility for a chain to have a zero gap requires that the following conditions are {\em both} satisfied: $\beta_{e}(j)=\beta$ and $\cos(k\theta) \, = \, -1$.
The second condition is satisfied for $k\theta = \pi$. 
Concerning the first condition, by  looking at Eq. \ref{eq:fock3} we see that $\beta_{e} (j)= \beta$ implies $\cos(\pi j/n)=1/2$, which is possible only if $n$ is a multiple of 3 (for $j=n/3$), leading to the important conclusion that
{\bf zigzag nanotubes are conductors if and only if $n$ is an integer multiple of 3}. This is a well know property of zigzag nanotubes. \cite{hamada1992new, saito1992electronic} 

The matrix elements of the Hamiltonian between symmetry-adapted orbitals are zero if the values of the transverse pseudo-momentum $j$ are different.
This means that each $j$ value is decoupled from the other ones, and corresponds formally to a 1-D system (either PBC or OBC, depending on the type of the boundary conditions of the nanotube).
In the following, it is important to consider the behaviour of the eigenvalues under the symmetry operation $\beta_e \rightarrow -\beta_e$.
In fact, by looking at Eq. \ref{eq:defepsilon}, it appears that the energy band is left unchanged if both $\beta_e$ and $\cos{k\theta}$ change their sign.
This means that an energy band is mapped onto itself under the operation $\beta_e \rightarrow -\beta_e$. 

Notice that the 1-D subsystems that compose the nanotube are not physical sub-chains in the nanotube itself.
Rather, each of them is composed of sites carrying a single function which are linear combinations, with different coefficients, of the radial $p$ AOs on the C atoms of a given set $\mu$ (all sharing the same $z$ coordinate).
Since all atoms involved in each linear combination have the same value of $z$, we can associate to each site this value of $z$. 
In this way, we can compute the TPS of each longitudinal chain as if it is a real dimerized chain with alternating bond lengths, $d$ and $d_{\rm e}$: $d$ (the usual Carbon-Carbon bond length in the tubes) for bonds having hopping integral $\beta$; and $d_{\rm e}=d/2$ for bonds having hopping $\beta_{\rm e}$.
However, it is important to stress the fact that the value of the effective hopping integral $\beta_e$ is not related to the value of the $z$ coordinates of the effective sites of the chains.
In fact, all Carbon-Carbon bonds in the nanotube are assumed to have the same length, and no  dimerization is present in the real system.
The value of $\beta_e$ depends on the difference in phases of connected atoms having their $\mu$ indices different by one in modulus. 
The value $d_e=d/2$, on the other hand, is a geometrical effect 
in the definition of the $z$ value of the sites.

The derivation here reported allows to strongly simplify the study of a zigzag
nanotube within the H\"uckel approximation, treating $n$ linear 1D systems with
$2m$ sites instead of a 2D system (folded along one dimension) with $2mn$
sites. Moreover, the full decoupling of the 1D systems makes simpler also the
calculation of the TPS and of the polarizability (in addition to the other one
electron properties), which are computed as the sum of the values of these
properties computed for the $n$ 1D systems. Concerning the conductivity
properties described at this level of theory and using an evocative image,
within an electric circuit a nanotube can be replaced with $n$ 1D ``cables``
(conveniently defined) working in parallel. In this scheme, the nanotube is a
conductor if at least one ``cable'' is a conductor.

It is worth highlighting that the analysis reported in this section is
performed within the H\"uckel approach. A short comment concerning the
comparison of the results here obtained with those expected with the use of the
exact Hamiltonian is in order at this point. This comparison is complex: from
one side one must consider that for all conjugated hydrocarbons the $\pi$ block
of the Fock matrix shows in general non-zero elements beyond the nearest
neighbor Carbon atoms and this leads to an interaction scheme in the symmetry
adapted basis defined in Eq. \ref{eq:site23} which is more complex than the one
here reported. On the other side, and perhaps more important, one must also
consider that the H\"uckel approach is based on a key assumption, the
hypothesis that the full MO manifold can be grouped in two subsets, those of
the $\pi$ and the $\sigma$ MOs. This partition is based on a symmetry
criterion, the behavior of the MOs with respect to a plane reflection
(symmetric for $\sigma$ MOs, antisymmetric for $\pi$ MOs), which implies that
the Fock matrix is block diagonal, with one block for the $\sigma$ manifold and
one for the $\pi$ manifold. In the canonical applications of H\"uckel theory
(as, for instance, linear polyenes and cyclic conjugated hydrocarbons), the
molecules are planar, this partition is exact and the block diagonal structure
of the Fock matrix is preserved also with the exact treatment. In the case of
the CNTs, the Carbon atoms are not on a plane and the partition of the MOs in
the here considered H\"uckel approach is based on a local pseudo symmetry
related to the plane tangent to the nanotube surface, as discussed in Sec.
\ref{sec:formal}. The absence of a real symmetry plane couples the $\sigma$ and
$\pi$ manifolds, the matrix of the exact Fock operator is no longer block
diagonal and this make very difficult the comparison of the H\"uckel approach
with the exact treatment.

\section{Computational Results}

In the present work, the computational study will be restricted to the case of periodic nanotubes. 
However, we stress the fact that within the scheme developed in Secs. \ref{sec:II}-\ref{sec:formal}, both periodic and open systems can be described.
In the preceding section, we have shown that the periodic zigzag nanotubes exhibit different types of electronic structures, depending on the number of hexagons around the tube.
In particular, the tubes have a vanishing energy gap at the Fermi level, and are therefore metallic, if the number of hexagons is a multiple of three, while they are insulators otherwise.
In any case, the energy gap is going to zero if the number of hexagons around the tube is going to $\infty$. 
Moreover, if the number of hexagons is even, one among the 1D chain in which the tube can be formally decomposed is totally dimerized,  that is, the value of $\beta_e$ for this 1D system is vanishing.
This implies that the orbital energies of this 1D chain can have only two values, $\beta$ and $-\beta$, independent of the length of the chain.
These different behaviours have a direct impact not only on the structure of the energy bands, but also on the position spreads and polarizabilities, both total and corresponding to the single chains.

For these reasons, we will present here numerical results for the four cases $n$=6, 7, 8, 9. In this way, the four different types of nanotubes will be presented, that is the combination of the metallic/insulating behavior (whether $n$ is or not a multiple of 3) with the presence/absence in the chain of non-interacting dimers (whether $n$ is or not a multiple of 2). We recall here that a chain composed by non-interacting dimers gives a constant/variable energy band, with constant/variable TPS and polarizability (depending on the fact that $n$ is even or odd).

\subsection{Orbital Energies}

\begin{figure}[ht]
\centering
\hspace{1cm}
\includegraphics[width=\textwidth]{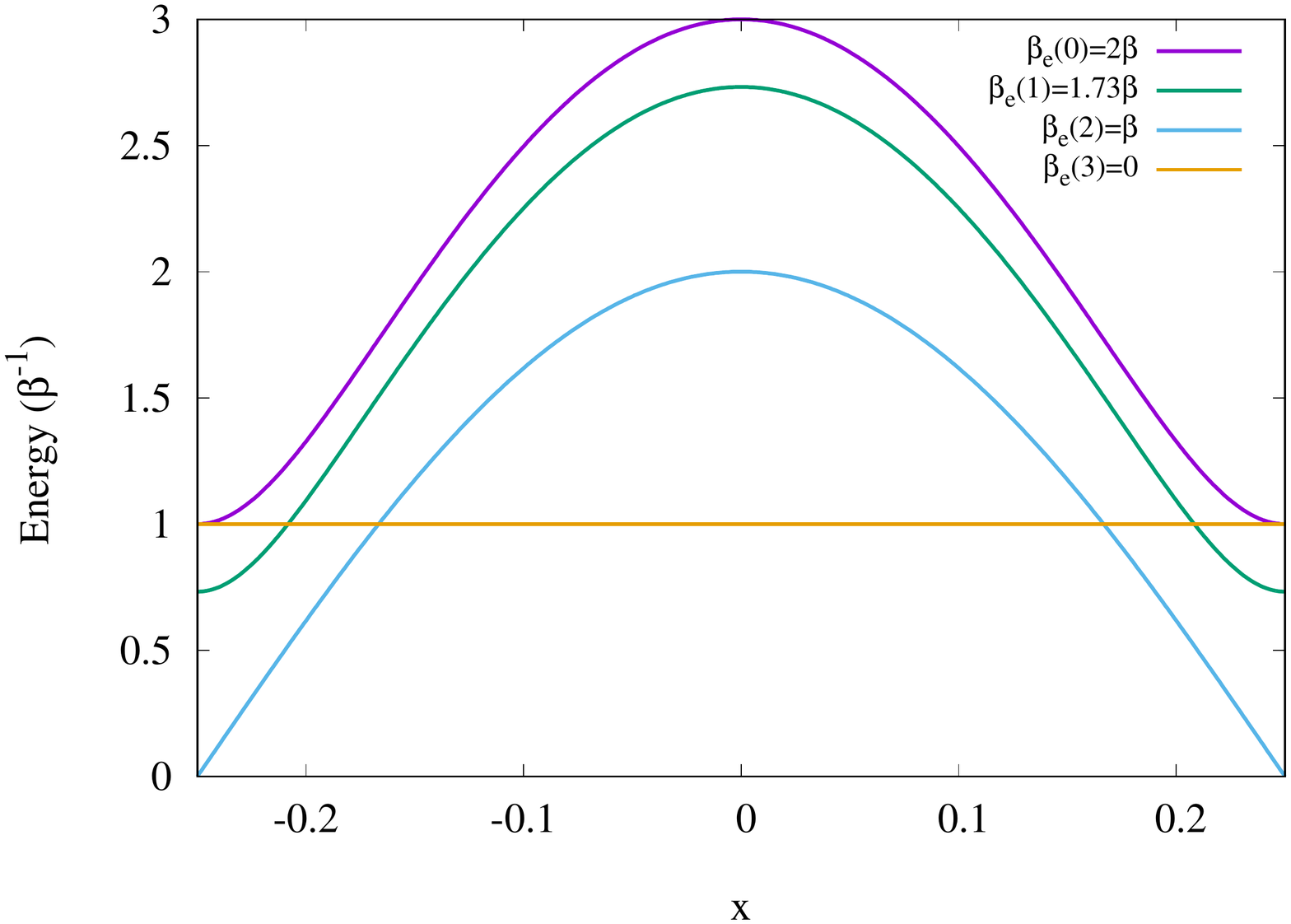}
\caption{Positive energy bands ($m\to\infty$) as a function of $x=k/m-0.5$ (see text for further details) of of the chains with $j=0,1,2,3$ of a zigzag nanotube with $n=6$ with PBC.}
\label{fig:PBChex6chains}
\end{figure}

\begin{figure}[ht]
\centering
\hspace{1cm}
\includegraphics[width=\textwidth]{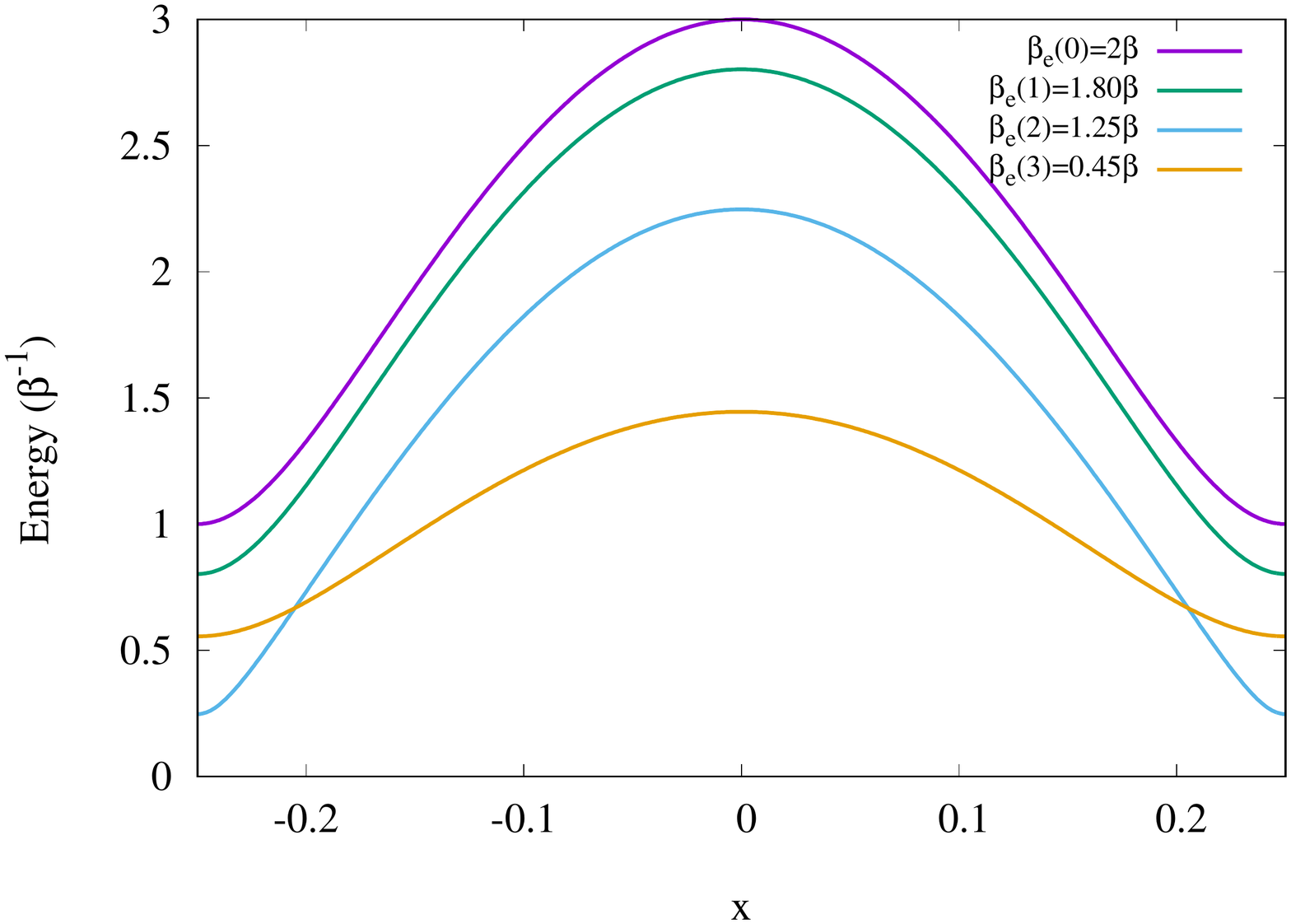}
\caption{Positive energy bands ($m\to\infty$) as a function of $x=k/m-0.5$ (see text for further details) of the chains with $j=0,1,2,3$ of a zigzag nanotube with $n=7$ with PBC.}
\label{fig:PBChex7chains}
\end{figure}

\begin{figure}[ht]
\centering
\hspace{1cm}
\includegraphics[width=\textwidth]{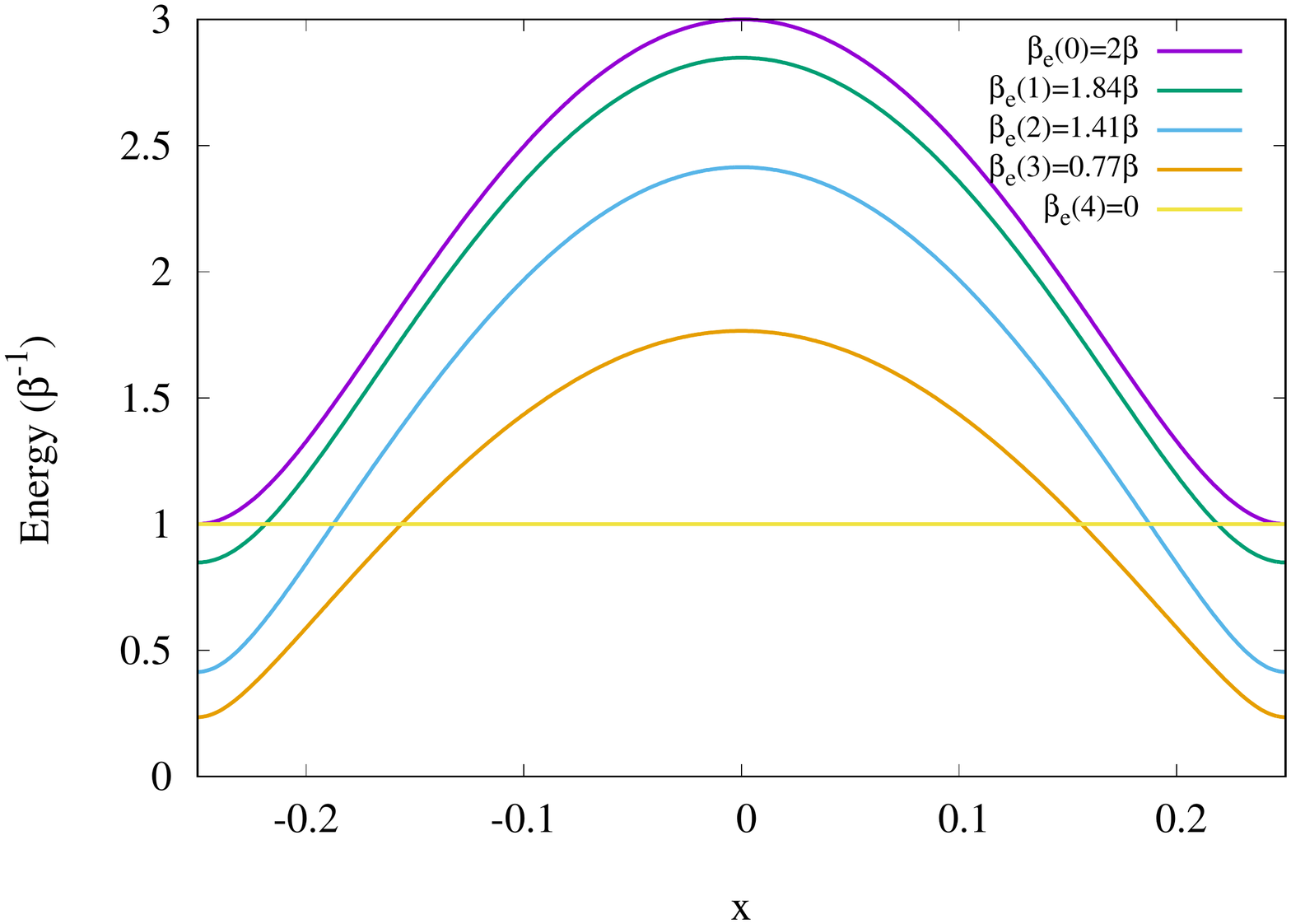}
\caption{Positive energy bands ($m\to\infty$) as a function of $x=k/m-0.5$ (see text for further details) of the chains with $j=0,1,2,3,4$ of a zigzag nanotube with $n=8$ with PBC.}
\label{fig:PBChex8chains}
\end{figure}

\begin{figure}[ht]
\centering
\hspace{1cm}
\includegraphics[width=\textwidth]{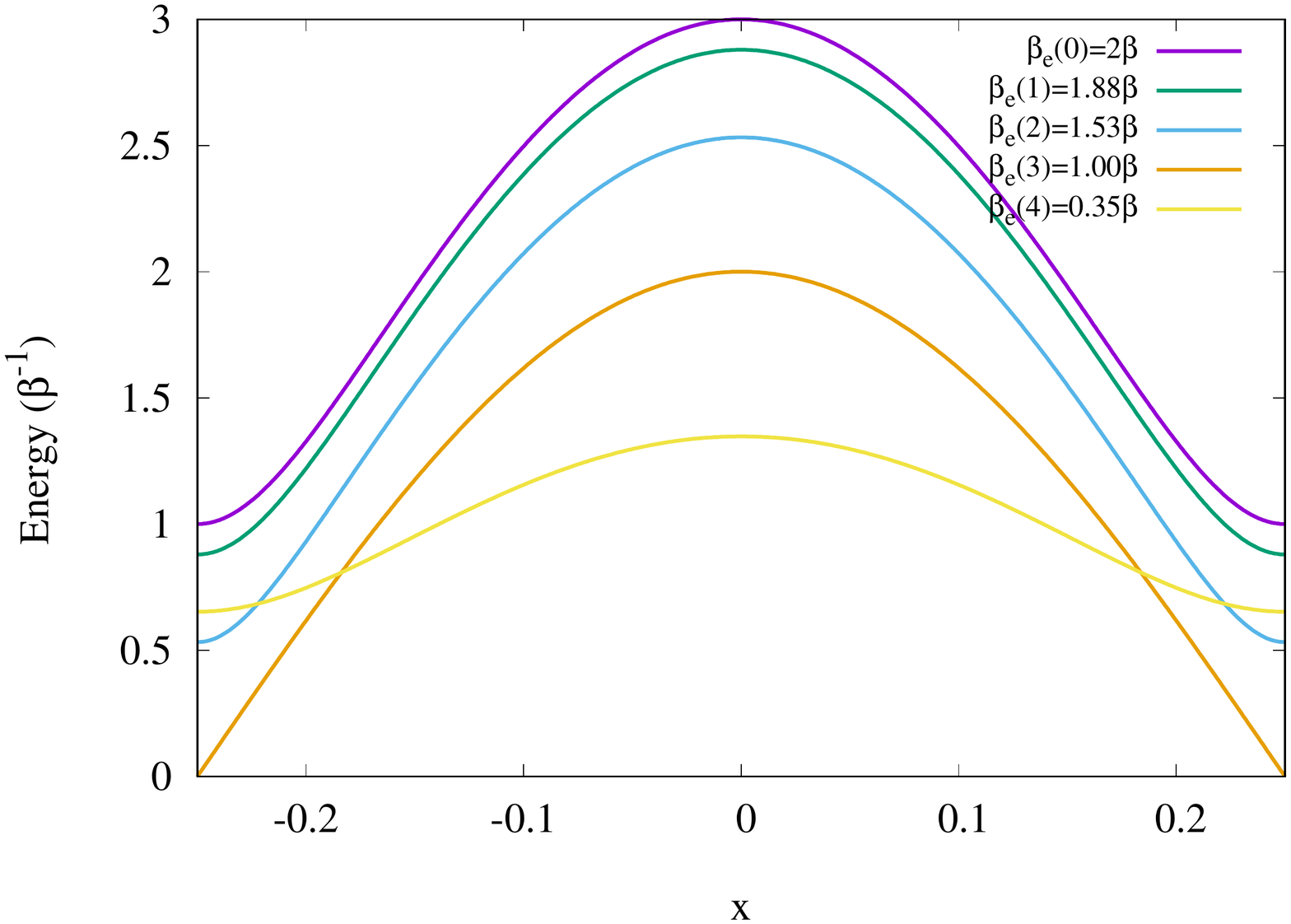}
\caption{Positive energy bands ($m\to\infty$) as a function of $x=k/m-0.5$ (see text for further details) of the chains with $j=0,1,2,3,4$ of a zigzag nanotube with $n=9$ with PBC.}
\label{fig:PBChex9chains}
\end{figure}

The one particle energies obtained by the strategy described above (see Eq. \ref{eq:defepsilon}), have been reported in Figs. 
\ref{fig:PBChex6chains}-\ref{fig:PBChex9chains} for a zigzag nanotube with $n=6$-$9$, respectively. 
The energies have been computed with $m=1000$ and are reported as a function of the discrete variable $x=k/m-0.5$.
The chosen value of $m$ can be considered as the limit $m\to\infty$ and within the scale of the figures, both the variable $x$ and the energies look continuous. 
Only the positive part of the spectrum of the one-electron Hamiltonian operator is reported and only in the $[m/4;3m/4]$ interval of $k$ (that is in the interval $[-0.25;0.25]$ for $x$).

\begin{figure}[ht]
\centering
\hspace{1cm}
\includegraphics[width=\textwidth]{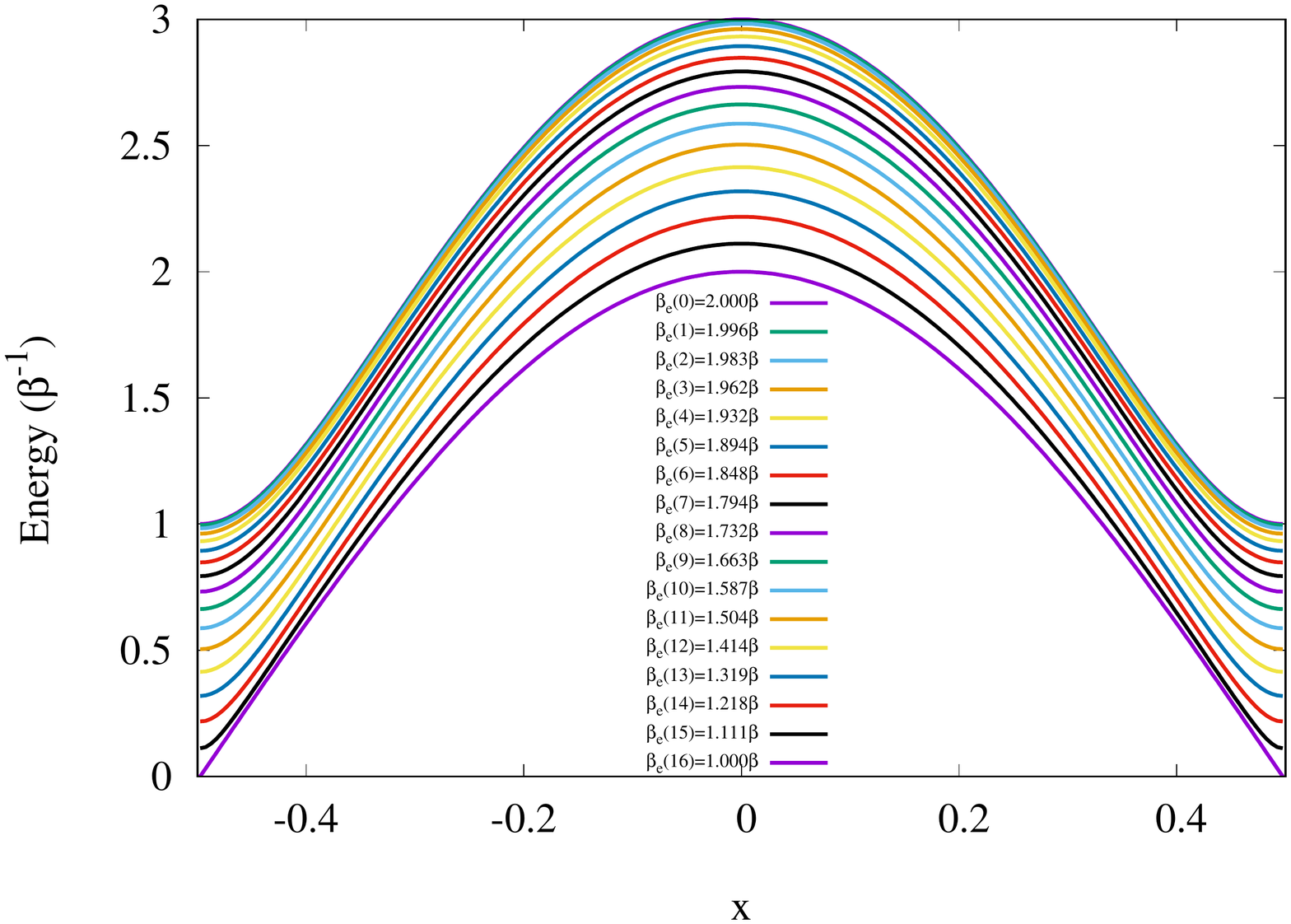}
\caption{Energy bands ($m\to\infty$) as a function of $x=k/m-0.5$ (see text for further details) of the chains with $j=0,...,16$ of a zigzag nanotube with $n=48$ with PBC.}
\label{fig:PBC_prima}
\end{figure}

\begin{figure}[ht]
\centering
\hspace{1cm}
\includegraphics[width=\textwidth]{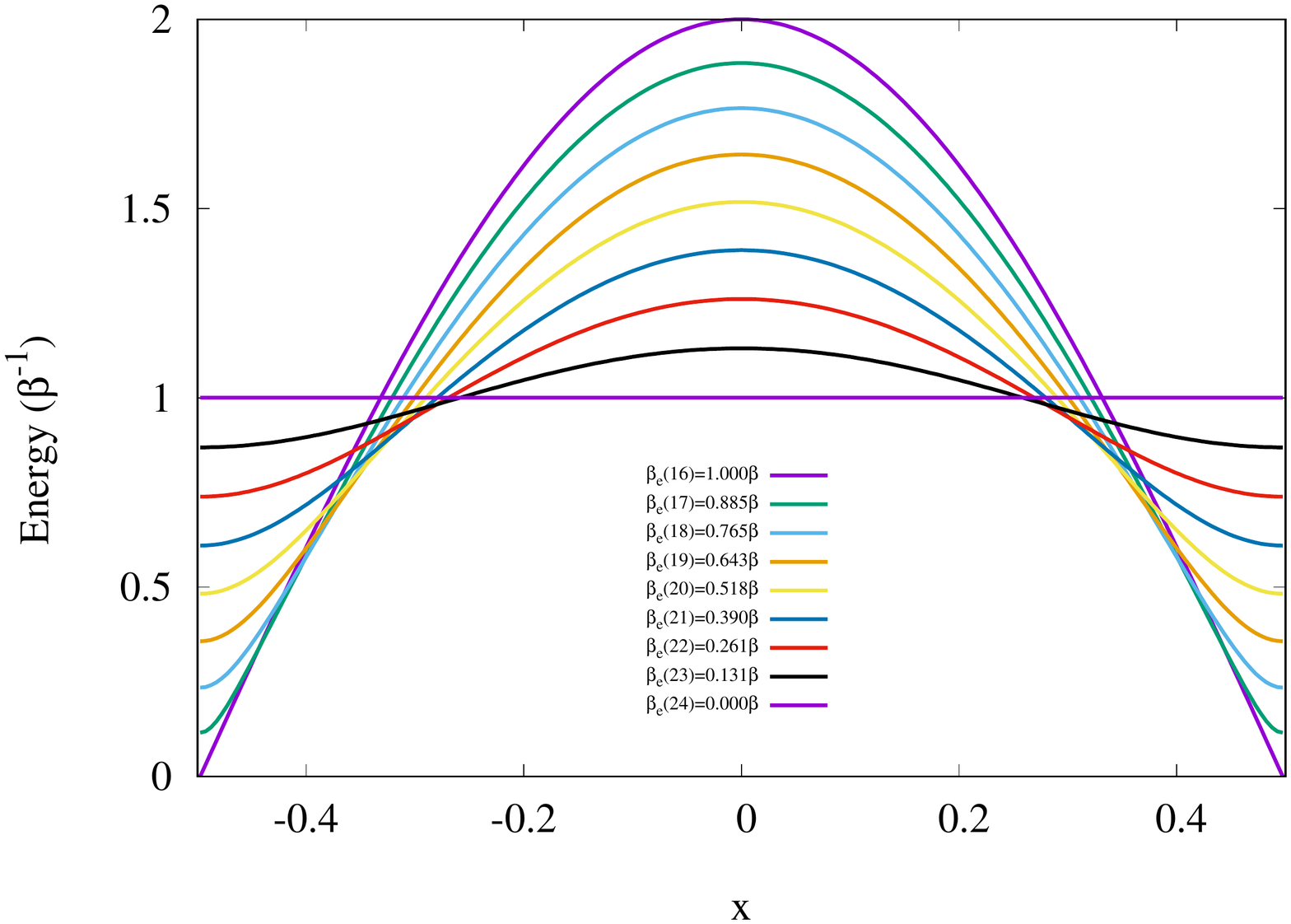}
\caption{Energy bands ($m\to\infty$) as a function of $x=k/m-0.5$ (see text for further details) of the chains with $j=16,...,25$ of a zigzag nanotube with $n=48$ with PBC.}
\label{fig:PBC_dopo}
\end{figure}

\begin{figure}[ht]
\centering
\hspace{1cm}
\includegraphics[width=\textwidth]{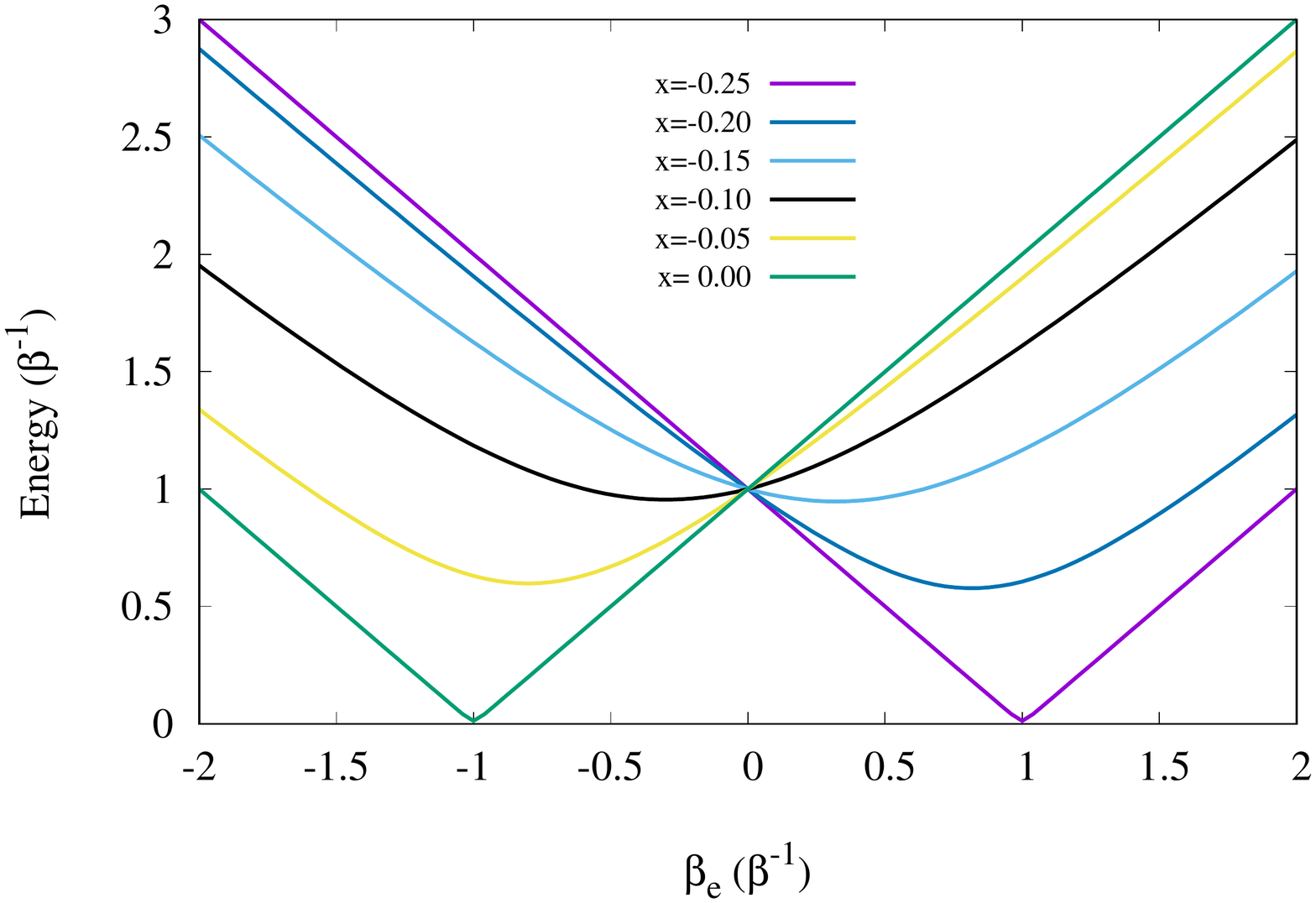}
\caption{Positive energy bands as a function of $\beta_e$ for
different values of $x=k/m-0.5$ (see text for further details) with PBC.}
\label{fig:Energy}
\end{figure}

For even values of $n$, the energy spectra contains two non-degenerate bands (those corresponding to $j=0$ and $j=n/2$), while all the other bands are doubly degenerate.
For odd values of $n$, on the other hand, all energy bands but the $j=0$ one are doubly degenerate.
For this reason, we can concentrate our discussion on the $j\leq n/2$ bands only.
All the bands characterized by a ``large'' value of the effective interaction $\beta_e$, \textit{i.e.} those having $\beta \le \beta_e \le 2 \beta$, do not cross each other and are comprised between the $\beta_e=\beta$ and the $\beta_e=2\beta$ bands (see Figure \ref{fig:PBC_prima}). The energies have been  computed for a large value of $n$ ($n=48$), which is both even and a multiple of 3.
In this way, the system is a metal and the constant band is present.
On the other hand, for ``small'' values of $\beta_e$, \textit{i.e.} when  $0 \le \beta_e \le \beta$, each band crosses all other bands, as shown in Figure \ref{fig:PBC_dopo}.
In Fig. \ref{fig:Energy} we report the positive energy bands (positive in $\beta$ units) as a function of the value of $\beta_e$ for different values of $x$. 
This figure graphically shows the considerations previously reported. 
For instance one sees that for $\beta_e=0$ (the linear chain is composed of independent ethylene units) the only possible energy in this part of the spectrum is $\beta$. 
Moreover, one promptly notes the equivalence of the spectrum for the transformation $\beta_e\rightarrow -\beta_e$. 
Finally, the metallic behavior of the chain with $\beta_e=\beta$ and of that with $\beta_e=-\beta$ is clear from the energy band gap closure at the Fermi level.

\subsection{Total-Position Spread}

\begin{figure}[ht]
\centering
\hspace{1cm}
\includegraphics[width=\textwidth]{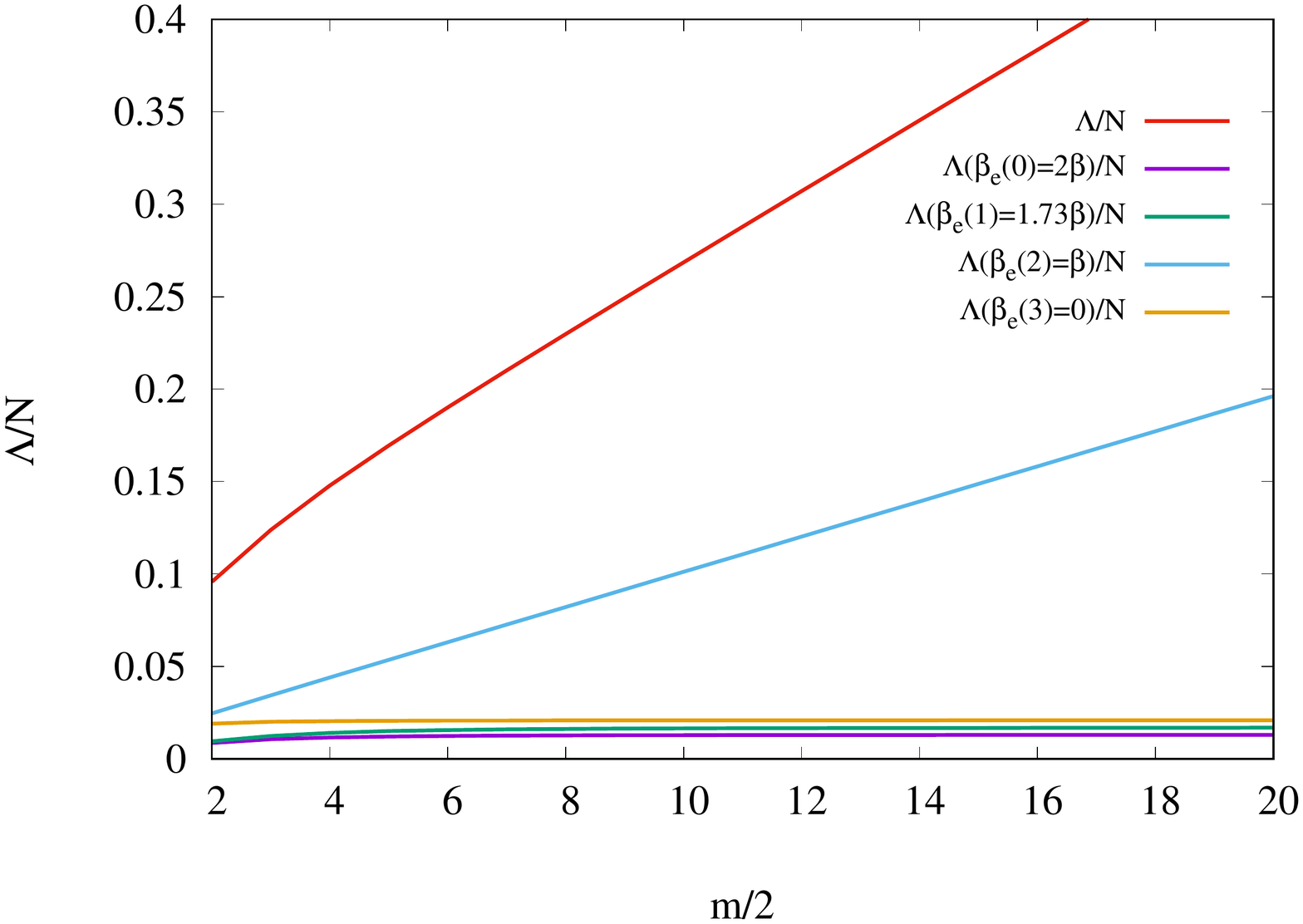}
\caption{TPS per electron of a zigzag nanotube with $n=6$ as a function of the length of the system (we use here the variable $m/2$ to stress 
than $m$ must be even).}
\label{fig:TPShex6}
\end{figure}

\begin{figure}[ht]
\centering
\hspace{1cm}
\includegraphics[width=\textwidth]{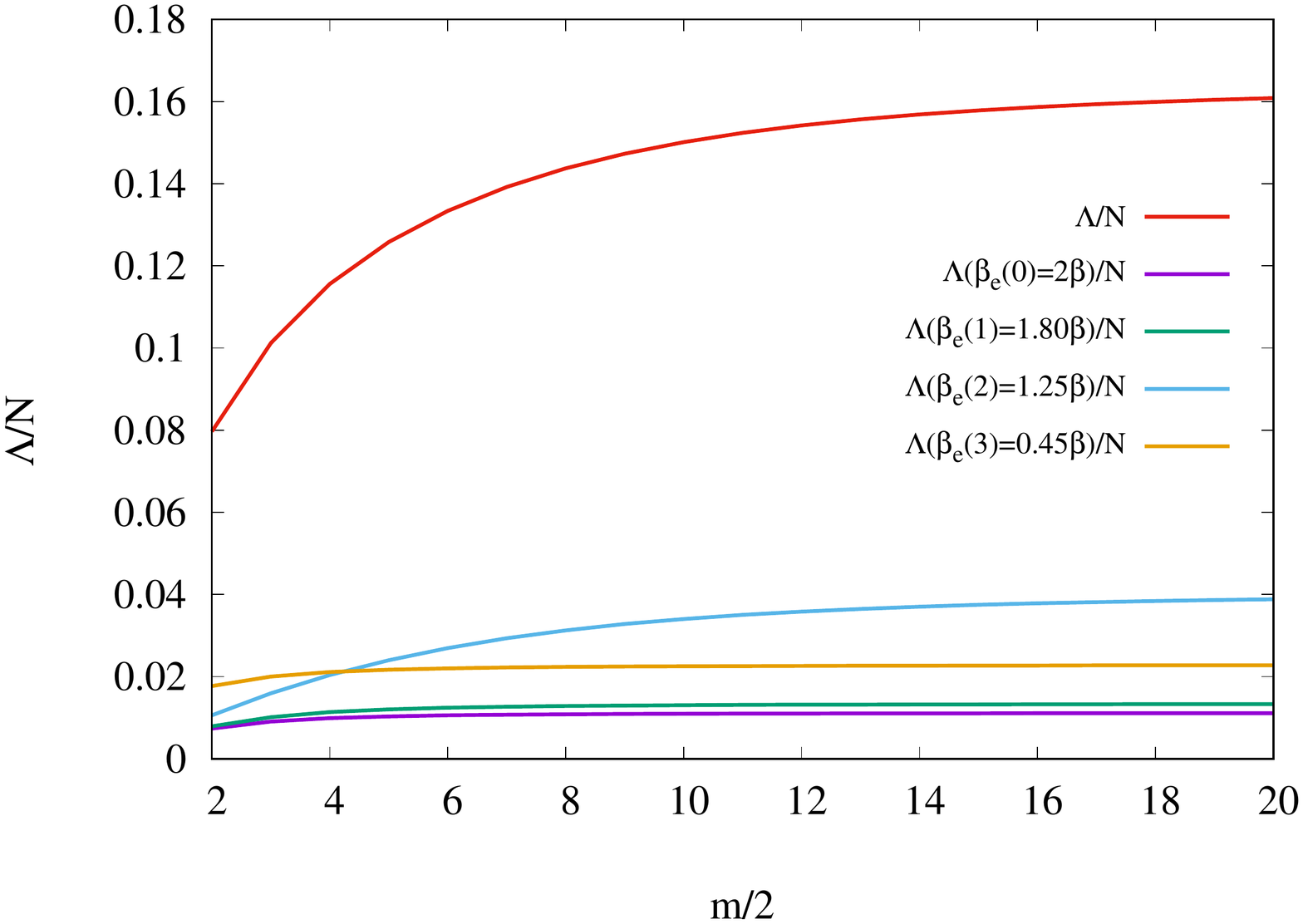}
\caption{TPS per electron of a zigzag nanotube with $n=7$ as a function of the length of the system (we use here the variable $m/2$ to stress 
than $m$ must be even).}
\label{fig:TPShex7}
\end{figure}

\begin{figure}[ht]
\centering
\hspace{1cm}
\includegraphics[width=\textwidth]{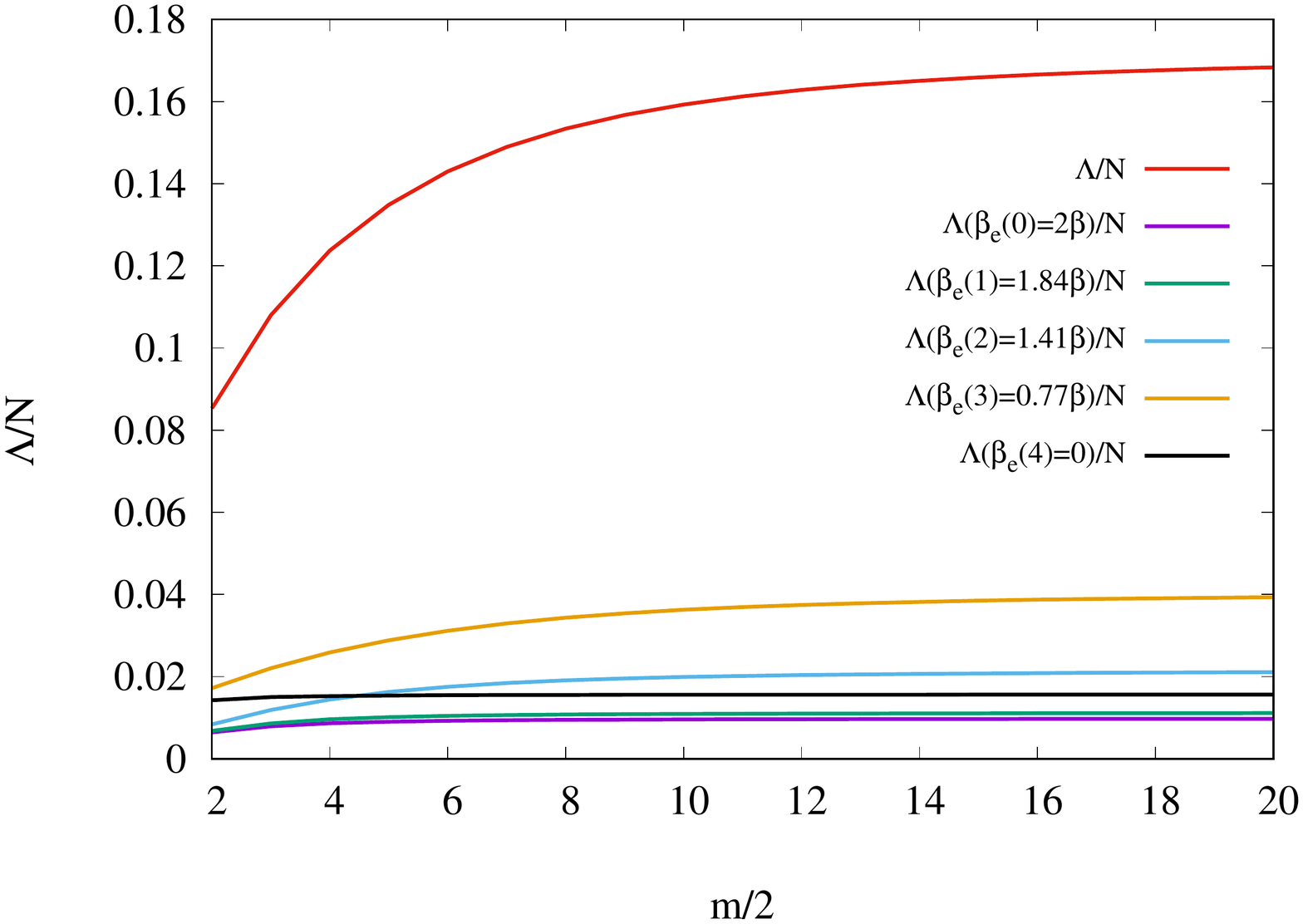}
\caption{TPS per electron of a zigzag nanotube with $n=8$ as a function of the length of the system (we use here the variable $m/2$ to stress 
than $m$ must be even).}
\label{fig:TPShex8}
\end{figure}

\begin{figure}[ht]
\centering
\hspace{1cm}
\includegraphics[width=\textwidth]{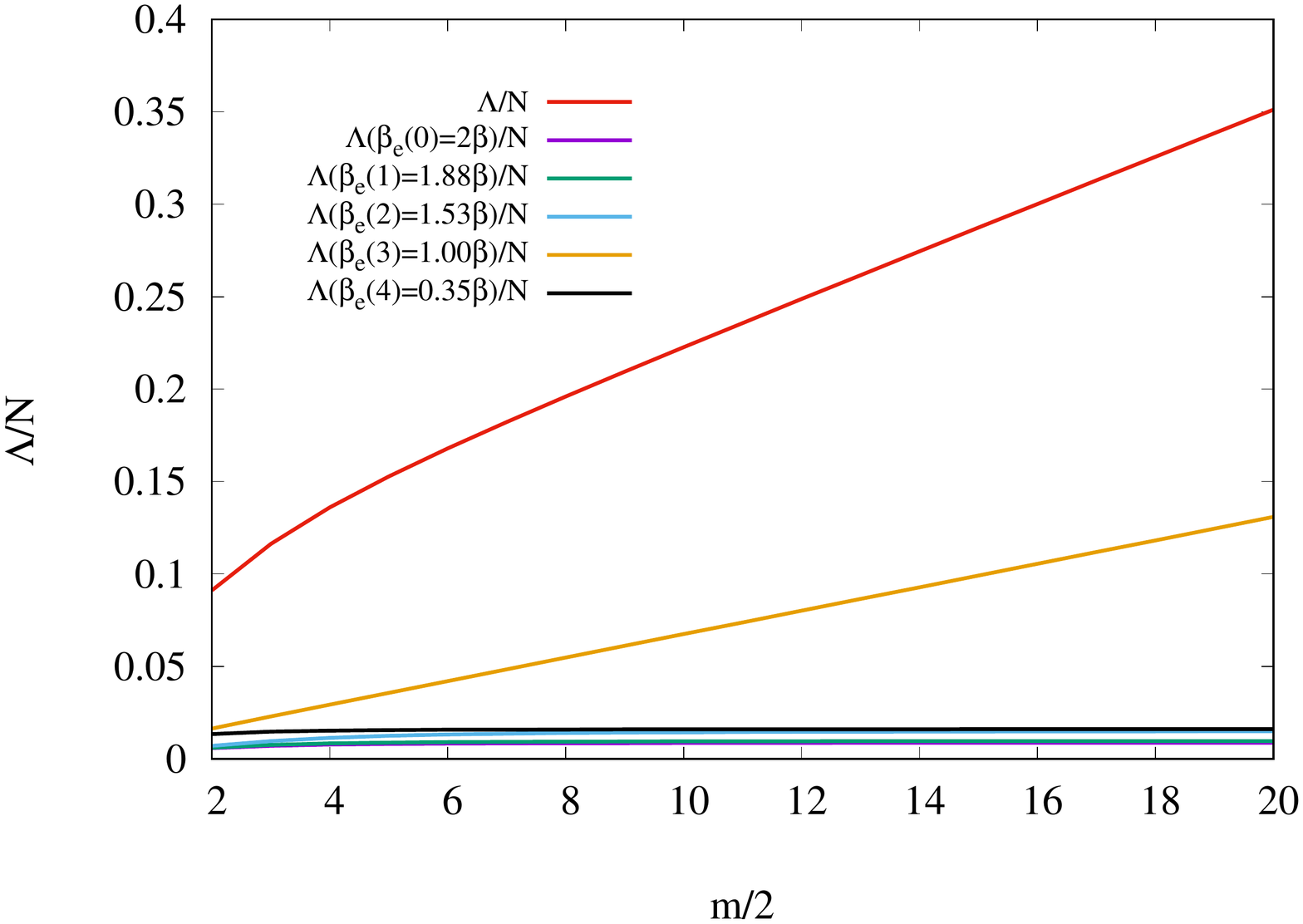}
\caption{TPS per electron of a zigzag nanotube with $n=9$ as a function of the length of the system (we use here the variable $m/2$ to stress 
than $m$ must be even). }
\label{fig:TPShex9}
\end{figure}

For the cases $n=6-9$, we show in Figures \ref{fig:TPShex6}-\ref{fig:TPShex9}, respectively, the TPS per electron, $\Lambda/N$ ($N$ being the total number of electrons, which in the H\"uckel approximation corresponds to the number of Carbon atoms, $2mn$, see Sec. \ref{sec:formal}), of the isolated virtual 1D chains and the corresponding total sum of these values over all the chains, that gives the TPS per electron of the entire system. 
As described in the Introduction Section, the asymptotic behaviour of the TPS per electron gives information about the conductivity of the system. 
The chains that correspond to either $\beta_e=2\beta$ or $\beta_e=0$ (if present) are not degenerate, while all the other ones are doubly degenerate and have the same TPS.
However, we represent only one of the degenerate pair in the figures. 

The metallic nanotubes (Figs. \ref{fig:TPShex6} and \ref{fig:TPShex9}, corresponding to the $n=6$ and $n=9$ cases, respectively) have two chains whose TPS per electron diverges linearly, as it is the case for metallic Annulenes.\cite{Angeli21}
These two 1D linear chains have both $\beta_e=\pm\beta$.
All the other chains show values of $\Lambda/N$ that saturate for large $N$.
For the $n=6$ and $n=9$ nanotubes the behavior of the total TPS per electron, therefore, is dominated by the contributions of the two metallic chains (metallic chains, if are there, are always doubly degenerate), and diverges linearly. On the other hand, for the nanotubes with $n=7$ and $=8$ (Figs. \ref{fig:TPShex7} and \ref{fig:TPShex8}, respectively) all 1D chains have a TSP per electrons that saturates for large $N$ and the total TPS is therefore finite at the thermodynamics limit.

\subsection{Polarizability}

\begin{figure}[ht]
\centering
\hspace{1cm}
\includegraphics[width=\textwidth]{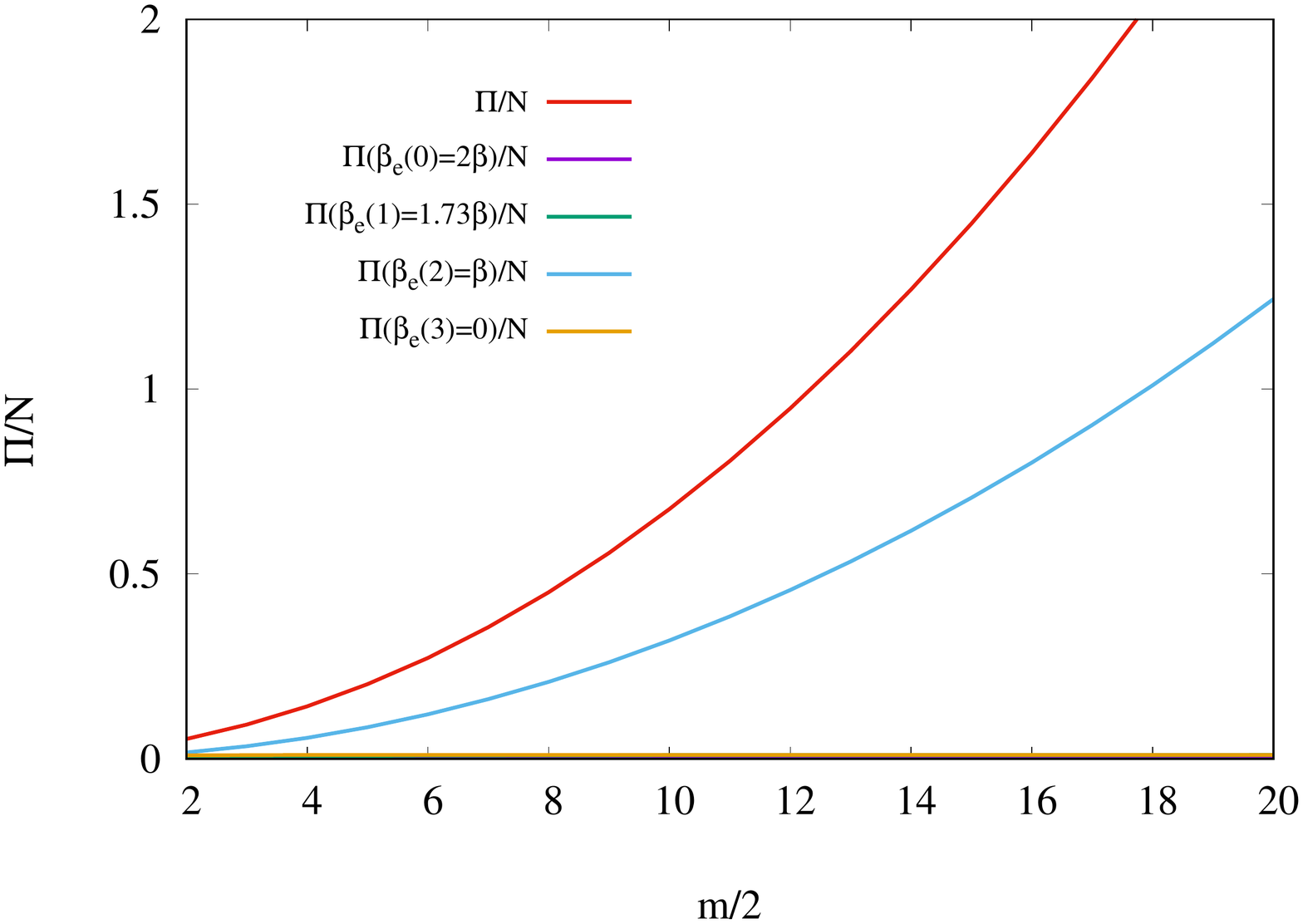}
\caption{Polarizability of a zigzag nanotube with $n=6$ as a function of the length of the system (we use here the variable $m/2$ to stress than $m$ must be even).}
\label{fig:Polhex6}
\end{figure}

\begin{figure}[ht]
\centering
\hspace{1cm}
\includegraphics[width=\textwidth]{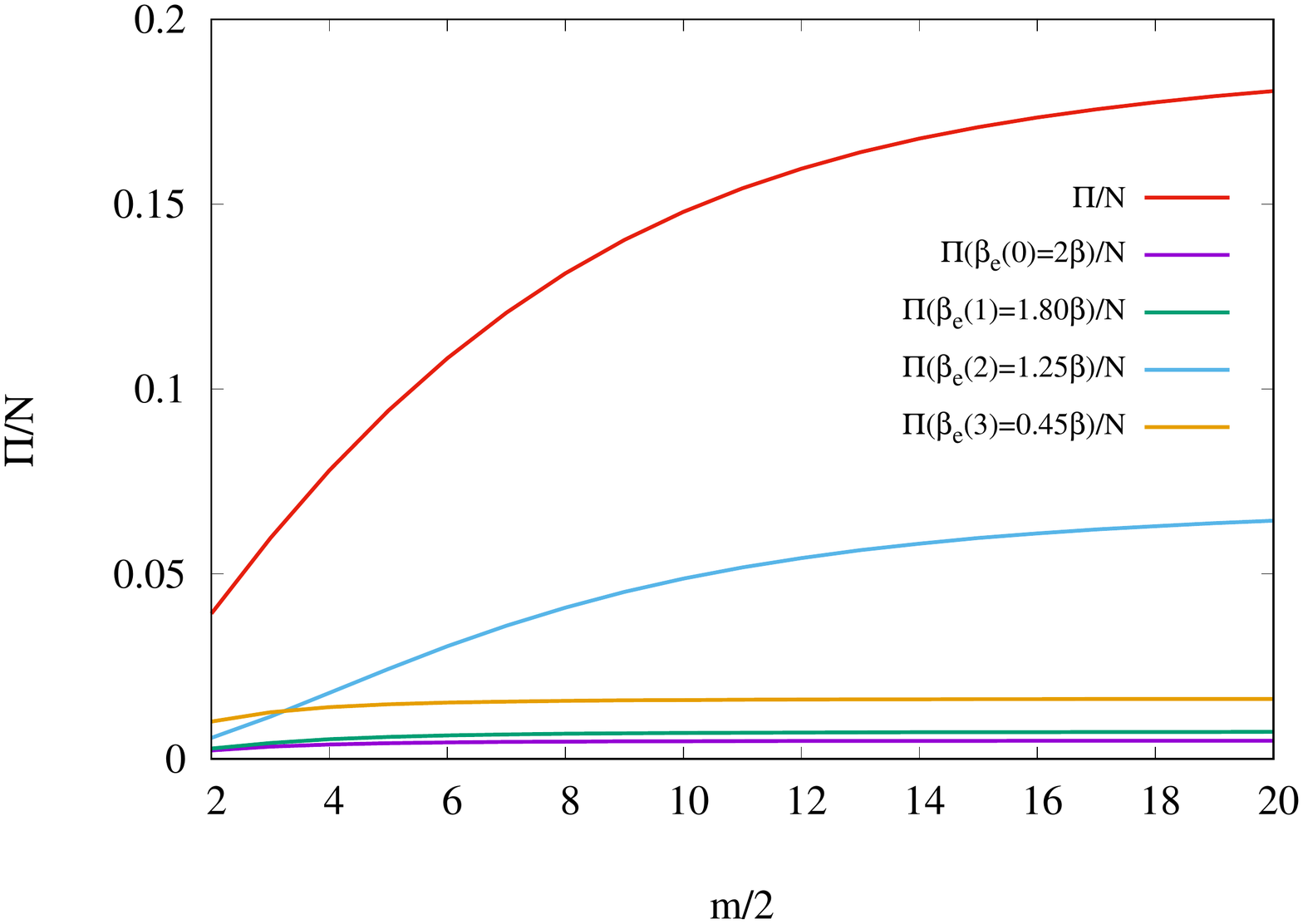}
\caption{Polarizability of a zigzag nanotube with $n=7$ as a function of the length of the system (we use here the variable $m/2$ to stress than $m$ must be even).}
\label{fig:Polhex7}
\end{figure}

\begin{figure}[ht]
\centering
\hspace{1cm}
\includegraphics[width=\textwidth]{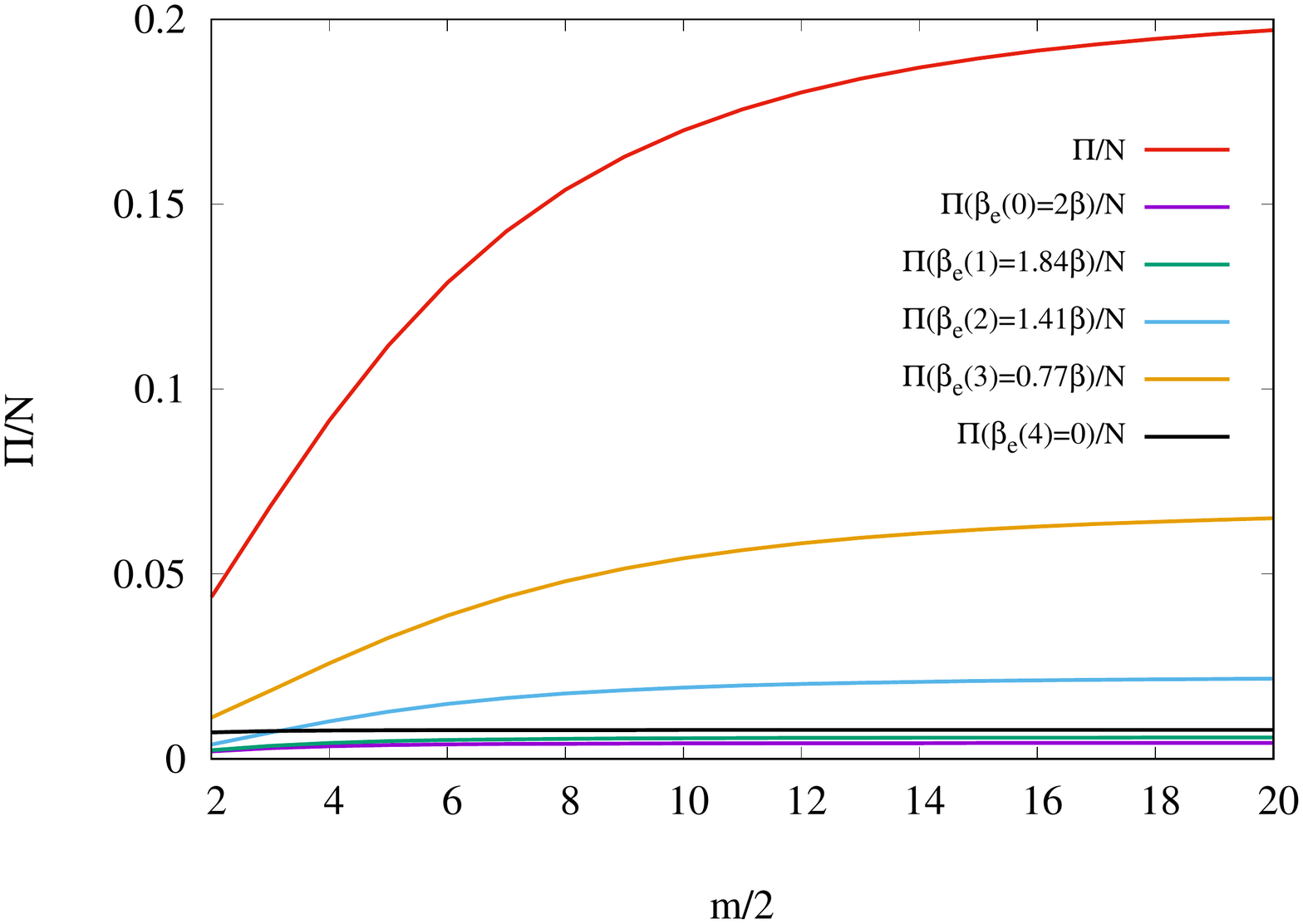}
\caption{Polarizability of a zigzag nanotube with $n=8$ as a function of the length of the system (we use here the variable $m/2$ to stress than $m$ must be even).}
\label{fig:Polhex8}
\end{figure}

\begin{figure}[ht]
\centering
\hspace{1cm}
\includegraphics[width=\textwidth]{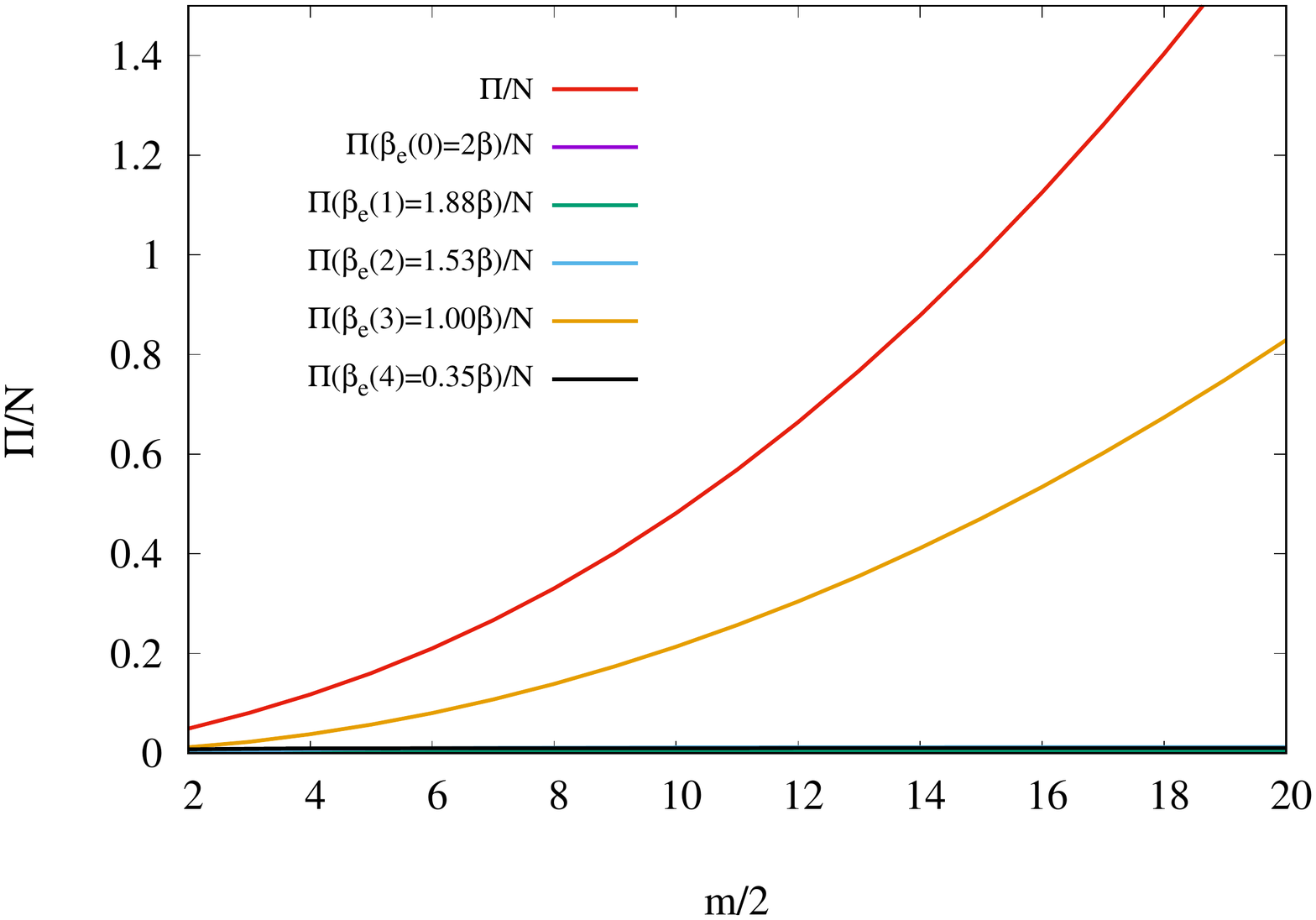}
\caption{Polarizability of a zigzag nanotube with $n=9$ as a function of the length of the system (we use here the variable $m/2$ to stress than $m$ must be even).}
\label{fig:Polhex9}
\end{figure}

In Figures \ref{fig:Polhex6}-\ref{fig:Polhex9} (for $n=6-9$, respectively) we show the Polarizability per electron, $\Pi/N$, of the isolated virtual chains and 
the corresponding total sum of these values over all the chains, that gives the Polarizability per electron of the entire system. 
The insulating-chain values of $\Pi/N$ have a behavior that is very much similar to the corresponding values of $\Lambda/N$. 
This can be seen, in particular, in Figs. \ref{fig:Polhex7} and \ref{fig:Polhex8}, with the single-chain values and the total sum that saturate to constant values.
The metallic chains (Figs. \ref{fig:Polhex6} and \ref{fig:Polhex9}), on the other hand, have a {\em quadratic} divergence for large values of $N$, and therefore the metallic contributions to $\Pi/N$ become overwhelmingly dominant for large values of $N$.
We notice again, as for the TPS case, that this behavior is consistent with that of Annulenes, described in Ref. [\onlinecite{Angeli21}].

\section{Conclusion}

We studied the electronic structure of zigzag Carbon Nanotubes in the 
H\"uckel approximation.
By using a suitable symmetry adaptation of the Atomic Orbitals, it is possible to decompose the system into a formal set of non-interacting chains (akin to linear polyenes, where the concept of the atoms bearing a $p$ AO in the polyene is replaced with those of the ``sites'' bearing a SALC in the virtual chains), whose treatment is particularly simple.
The energy spectrum of these nanotubes has been investigated, by studying the individual behaviour of the 1D chains and thus taking advantage of this decomposition, recovering the results known in the literature: systems having a number of hexagons around the tube that is a multiple of three are gapless 
systems (metals), while they have a non-zero gap at the Fermi level for the other cases (insulators).
However, for conductivity our approach based on the TPS goes well beyond the simple analysis on the energy gap, which is inherently a one-electron description.
In fact, the divergence of the per-electron TPS can be used for one-electron and many-electron Hamiltonians.
Moreover, we notice that the symmetry adaptation presented here is possible for both finite open nanotubes and topologically closed nanotubes, most suited to the treatment of an infinite system.
Although in this work we limited our investigation to closed structures, the possibility of treating open systems is interesting for investigating short open nanotubes.

These results have been confirmed by computing the per-electron longitudinal position spread and the polarizability of the systems.
It has been shown that the position spread diverges linearly in the case of metallic systems as a function of the nanotube length, while the divergence of the polarizability is quadratic in the same case. 
Both quantities, on the other hand, saturate to constant values in the case of insulators.
Most importantly, we have shown that the periodic position operator that we recently introduced in Refs. [\onlinecite{Valenca, Angeli21,evangelisti_PhysRevB.105.235201}], is suitable to compute the position spread and the polarizability, 
quantities which cannot be obtained in periodic systems  using the ``ill defined'' standard position operator.
In future works, we plan to generalize our investigation to armchair nanotubes, as well as tubes having a non-zero helicity around their axe. 
Moreover, by considering the limit of very long and wide nanotubes, a surface of Graphene is obtained, so our approach will be able to compute position spreads and polarizabilities of this much studied and interesting system.

\section*{Acknowledgements}

We thank the French ``Agence Nationale de la Recherche (ANR)'' for financial
support (Grant Agreements No. ANR-19-CE30-0011 and ANR-22-CE29-0001). This work
has been (partially) supported through the EUR grant NanoX n$^\circ$
ANR-17-EURE-0009 in the framework of the ``Programme des Investissements
d'Avenir''.

\section*{Appendix A: An alternative form of Bloch Theorem}

Let us consider a one-electron Hamiltonian in a 1-D periodic system, and let us assume
that the system is composed of $n$ identical blocks.
The matrix representing the Hamiltonian will be of the form
\begin{equation}
\hspace{-16mm}{\bf H} \; = \; 
\left\|
\begin{array}{lllllllllllllccccccccc}
{\bf H}_0  &  {\bf H}_1  &  {\bf H}_2  &  ...  &  {\bf H}_{n-2}  &  {\bf H}_{n-1} \\
{\bf H}_1^+  &  {\bf H}_0  &  {\bf H}_1  &  ...  &  {\bf H}_{n-3}  &  {\bf H}_{n-2} \\
... \\
{\bf H}_{n-1}^+  &  {\bf H}_{n-2}^+  &  {\bf H}_{n-3}^+  &  ...  &  {\bf H}_{1}^+  &  {\bf H}_{0} \\
\tag{A1}
\end{array}
\right\| \hspace{6mm} .\hspace{10mm} 
\end{equation}

We seek an eigenvector of the form
\begin{equation}
\hspace{-16mm}{\bf \Psi} \; = \; 
\left\|
\begin{array}{lllllllllllllccccccccc}
{\bf \psi}  \\
e^{ik\theta} \, {\bf \psi}  \\
...  \\
e^{i(n-1)k\theta} \, {\bf \psi}  \\
\end{array}
\right\| \hspace{10mm} 
\tag{A2}
\end{equation}
where $\theta=2\pi/n$.
It is straightforward to verify that $\psi \, = \, \psi_k$ must be an eigensolution of the of the Effective Hamiltonian
\begin{equation}
{\bf \bar H}(k) \; = \; \sum_{l=0}^{n-1} \, e^{ikl\theta} \, {\bf H}_l \hspace{10mm} k=0,\ldots, n-1
\tag{A3}
\end{equation}
in such a way that the eigenequations for ${\bf \bar H}(k)$ will be written as
\begin{equation}
{\bf \bar H}(k) \, |\psi_{k,\mu}\rangle  \; = \; \epsilon_{k,\mu} |\psi_{k,\mu}\rangle \hspace{10mm} k=0,\ldots, n-1
\tag{A4}
\end{equation}

\section*{Data Availability}

The data that supports the findings of this study are available within the article.

\end{document}